\documentclass{article}
\usepackage{epsfig}
\usepackage{subfigure}
\usepackage{caption}
\usepackage{subfig}
\usepackage{url}
\usepackage{cite}
\usepackage{authblk}

\title{Spatio-Temporal Analysis of Topic Popularity in Twitter}
\author[1]{Sebastien Ardon}
\author[2]{Amitabha Bagchi}
\author[1]{Anirban Mahanti}
\author[2]{Amit Ruhela}
\author[2]{Aaditeshwar Seth}
\author[2]{Rudra Mohan Tripathy}
\author[1]{Sipat Triukose}

\affil[1]{NICTA, Australia, Email
  \{sebastian.ardon,anirban.mahanti,sipat.triukose\}@nicta.com.au}
\affil[2]{Department of CS\&E, IIT Delhi, Email \{bagchi,aruhela,aseth,tripathy\}@cse.iitd.ernet.in}

\begin{document}

\maketitle

\begin{abstract}

We present the first comprehensive characterization of the diffusion
of ideas on Twitter, studying more than 4000 topics that
include both popular and less popular topics. On a data set containing
approximately 10 million users and a comprehensive scraping of all the
tweets posted by these users between June 2009 and August 2009
(approximately 200 million tweets), we perform a rigorous temporal and
spatial analysis, investigating the time-evolving properties of the
subgraphs formed by the users discussing each topic. We focus on two
different notions of the spatial: the network topology formed by
follower-following links on Twitter, and the geospatial location of
the users. We investigate the effect of initiators on the popularity
of topics and find that users with a high number of followers have a
strong impact on popularity. We deduce that topics become popular when disjoint clusters
of users discussing them begin to merge and form one giant component
that grows to cover a significant fraction of the network. Our geospatial analysis shows
 that highly popular topics are those that cross regional boundaries aggressively.

\end{abstract}




\section{Introduction}
\label{sec:intro}

We live in the era of Twitter. From the shenanigans of pop stars and
actors to enduring political transformations, everything is being
transacted on microblogging services. Nonetheless, fundamental questions remain
unanswered. We know, for instance, that discussions around certain
topics ``go viral'' whereas other topics die an early death. The
network propagates some ideas, and some make no headway. In view of
the enormous influence of online social networks (OSN), understanding the 
mechanics of these systems is critical. To
characterize the properties of popular and non-popular topics is of
surpassing importance to our understanding of how these complex
networks are shaping our world. 

In this paper we present a large-scale measurement study that attempts
to describe and explain the processes that animate microblogging
services. We study a large set of popular and non-popular topics
derived from a comprehensive data set of tweets and user information
taken from Twitter. A key strength of our study is that we observe
both popular and not-so-popular topics. This allows us to hypothesize about
the temporal and spatial behavior of popular topics and support our
hypotheses by showing that non-popular topics display contrary behavior.

Note that we use the more general term {\em popular} rather than the
more specific term {\em viral.} This is to make a clear distinction
between those topics that achieve popularity because of processes and
situations that lie {\em outside} the network and those whose
popularity can be attributed to the dynamics that take place within
the network. We reserve the term viral for those topics whose vast
popularity is a product of the social network's internal
dynamics. These topics could not have gained popularity in the pre-OSN
era unless traditional news media decided to promote them. Our study
does not focus on these kinds of topics in particular because we intend
to study the entire ecosystem.


Our work emphasizes the structural aspects of topic spread. We give
the semantic aspect its due importance in the process of topic
identification and then proceed to study the fundamental temporal and
spatial aspects of the spread of topics. In particular, we study topic movement 
over two interrelated spatial dimensions: the topology of the Twitter
network as formed by ``follower'' and ``following'' relationships, and
the geospatial embedding of that network in the map of the world.

Our study spans several aspects of spatial diffusion, but our
primary focus is on characterizing the temporal and spatial
underpinnings of popularity. We focus on three important aspects 
as described in the sequel. First, in Section~\ref{sec:initiators}, we study 
how topic initiators influence popularity of the topic, and make
the following observations:
\begin{description}
\item{\bf Hypothesis 1.} Twitter is a partially democratic medium in
  the sense that popular topics are generally started by users with
  high numbers of followers (we call them celebrities); however, for a topic
  to become popular it must be taken up by non-celebrity users. 

\item{\bf Corollary 1a.} Regions with large user bases or with large
  number of heavily followed celebrities and news sources dominate
  Twitter.  
\end{description}

Second, in Section~\ref{sec:topology}, we study the effect of topology and the
dynamics of topic spread on popularity. The primary objects of study
to this end are the subnetworks formed by users discussing each
topic. While it is known that the Twitter network, like most large
OSN, contains a giant connected component, a key 
finding is that the subgraph of users talking about a popular topic on
a particular days always contains a giant connected component
containing most of the nodes (users) of the subgraph, whereas the
subgraphs of non-popular topics tend to be highly disconnected. To 
summarize, we make the following observations:
\begin{description}
\item{\bf Hypothesis 2.} Most of the people talking about a popular
  topic on a given day tend to form a large connected subgraph (giant component)
  while unpopular topics are discussed in disconnected
  clusters. 
\item{\bf Hypothesis 2a.} The giant component forms when many tightly
  clustered sets of users discussing the topic merge. 
\end{description}

Finally, we study the impact of geography on popularity by partitioning
the Twitter network according to regional divisions and studying the
behavior of popular and non-popular topics.
\begin{description}
\item{\bf Hypothesis 3.} Popular topics cross regional boundaries
  while unpopular topics stay within them.
\end{description}
The evidence for this observation is presented in
Section~\ref{sec:geography}. 

Apart from the highlights mentioned above, we review
 related work in Section~\ref{sec:related}. We
describe the various methodological issues that needed to be
surmounted to perform our study in Section~\ref{sec:methodology}. 
Section~\ref{sec:conclusions} concludes the paper with a discussion of the
implications of our observations on different aspect of the OSN
sphere.

\section{Background and Related Work}
\label{sec:related}

Leskovec, Backstrom and Kleinberg's seminal work on the evolution of
topics in the news sphere was the starting point for this
paper~\cite{Leskovec_2009}. They studied how the growth of one topic
affects the growth of other topics in the blogosphere. They identified
and tracked a small number of popular threads, and showed that the
growth of the number of posts on a thread negatively impacts the
growth of other threads. The basic question that arose on reading that
work was this: Can the nuances of the temporal evolution of topics be
explained by a more thorough study of their spatial evolution? Working
with a data set taken from Twitter we were able to extract the high
level of structural and geographical information about the actors of
the process that has allowed us to answer this question in the
affirmative. This allows us to challenge the line of research that
studies only the temporal evolution of topics~\cite{Yang_2011},
or seeks to explain this evolution on the basis of
content~\cite{wu-icwsm:2011}.

Following the paper cited above there has been more interest in
understanding how information and ideas propagate on OSNs. A
pioneering study on these phenomena on Twitter was conducted by Kwak
et. al.~\cite{Kwak_2010} where several aspects of topic diffusion were
studied. Of particular relevance to our work was their study on the
topological properties of retweet trees. Since our data set is built
on the data set they used (cf. Section~\ref{sec:methodology}) for
details), our work can easily be compared. Our major contribution is
that we work with a more general notion of a topic and that we work
with an ecosystem of topics. Also our work views the diffusion of
topics through the lens of what we call ``topic graphs'' (cf.
Section~\ref{sec:topology}), that are a significant
generalization of retweet trees. Retweet cascades have also been
studied specifically for the case of tweets with URLs in them by
Galuba et. al.~\cite{galuba-wosn:2010} and by Rodrigues
et. al.~\cite{rodrigues-imc:2011}.  There is a line of work that seeks
to uncover the structural processes behind topic diffusion by studying
cascade models (e.g. Ghosh and Lerman~\cite{ghosh-wsdm:2011}, Sadikov
et. al.~\cite{sadikov-wsdm:2011}) but we feel this is a limited view
of the effect of topology and try to view the network structure in a
more complex way.

In another work relevant to ours, Sousa el al.~\cite{Sousa_2010}
investigated whether user interactions on Twitter are based on social
ties or on topics, by tracking replies and message exchange on
Twitter; their study is focused on only three topics namely sports,
religion, and politics.  More recently, Romero et
al.~\cite{Romero_2011} studied topic diffusion mechanisms on Twitter
by focusing on topics identifiable by hashtags.  They study the
probability of a topic adoption based on repeated exposure, and
provide quantitative evidence of a contagion phenomenon made more
complex than normal studies of virus-like phenomena by the existence
of multiple topics, and briefly report on the graph structure of topic
networks. One major limitation of this work we found is that only a
very small fraction (approx. 10\%) of tweets are tagged with hashtags
(see Table~\ref{table:dataset} in Section~\ref{sec:methodology}). Our
methodology of using a Natural Language Processor
(OpenCalais~\cite{opencalais_url}) allows us to study topic diffusion
on a much larger scale than in this work since our topic choices are
not limited to hashtags.  On the geographical front, Yardi et
al.~\cite{Yardi_2010} examine information spread along the social
network and across geographic regions by analyzing tweets related to
two specific events happening at two different geographic
locations. As an aside we mention that Krishnamurthy
et. al. characterized the geographical properties of the Twitter user
base in 2008~\cite{krishnamurthy-wosn:2008}.

On a more general level, we note that it is implicitly assumed that
the attention of users on a platform like Twitter is elastic but
bounded (see e.g.~\cite{lotan-socialflow:2011}) and hence the
diffusion process is essentially a competitive one, even if it is not
explicitly adversarial. The study of competitive diffusion has largely
revolved around the application domain of viral marketing where there
is competition between different products~\cite{Bharathi_2007,
  Carnes_2007, Tomochi_2005, iribarren-phyreve:2011}. Budak et
al. \cite{Budak_2011} consider the problem of diffusion of
mis-information, where opposing ideas are competing and propagating in
a social network. The study of processes by which rumor spread may be
combated is another example of competitive
diffusion~\cite{Tripathy_2010}. Our work provides an important input
into this area of study, articulating the properties of a complex
system that requires extensive study to model correctly and
comprehensively.

\section{Methodology}
\label{sec:methodology}

\subsection{Data Set Description}
\label{sec:dataset}

We used a portion of the `tweet7' data set crawled by Yang et
al.~\cite{Yang_2011}. This data set contains 467 million tweets,
collected over a period of seven months, from June to December, 2009.
The tweets emanated from over 17 million users and are estimated to
constitute about 20-30\% of all tweets posted during that time period.
For our analysis, we used the first three month's tweets of this data
set.

\begin{table}[!t]
\renewcommand{\arraystretch}{1.3}
\centering
\begin{tabular}{|l|r|}
\hline
Tweets & 196,985,580\\
\hline
Users & 9,801,062\\
\hline
Hashtags & 1,341,733\\
\hline
Tweets with Hashtags  & 19,043,104\\
\hline
Retweets & 15,126,588\\
\hline
Tweets with URLs & 54,443,857\\
\hline
Direct (@) Tweets & 41,951,786\\
\hline
\end{tabular}
\caption{Data set summary.}
\label{table:dataset}
\end{table} 

The high-level description of the data we used is shown in
Table~\ref{table:dataset}.  As shown, out of approximately $200$
millions tweets, only approximately $20$ millions tweets contain a
hashtag. Next, we augmented this data set with network information, in
particular, the Twitter follower-following relationship between users,
using a separate data set collected from Twitter during the same time
period by Kwak et al.~\cite{Kwak_2010}. This allowed us to construct a
directed graph of Twitter users, where a node represents a Twitter
user and an edge between two nodes $u$, $v$, represented $(u
\rightarrow v)$, represents that $v$ is a follower of $u$. This
captures the fact that if $v$ follows $u$ then $u$'s tweets are
visible directly on $v$'s timeline.

\begin{figure*}[t]
\centering
\begin{tabular}{ccc}
\begin{minipage}{5.3cm}
\epsfig{figure=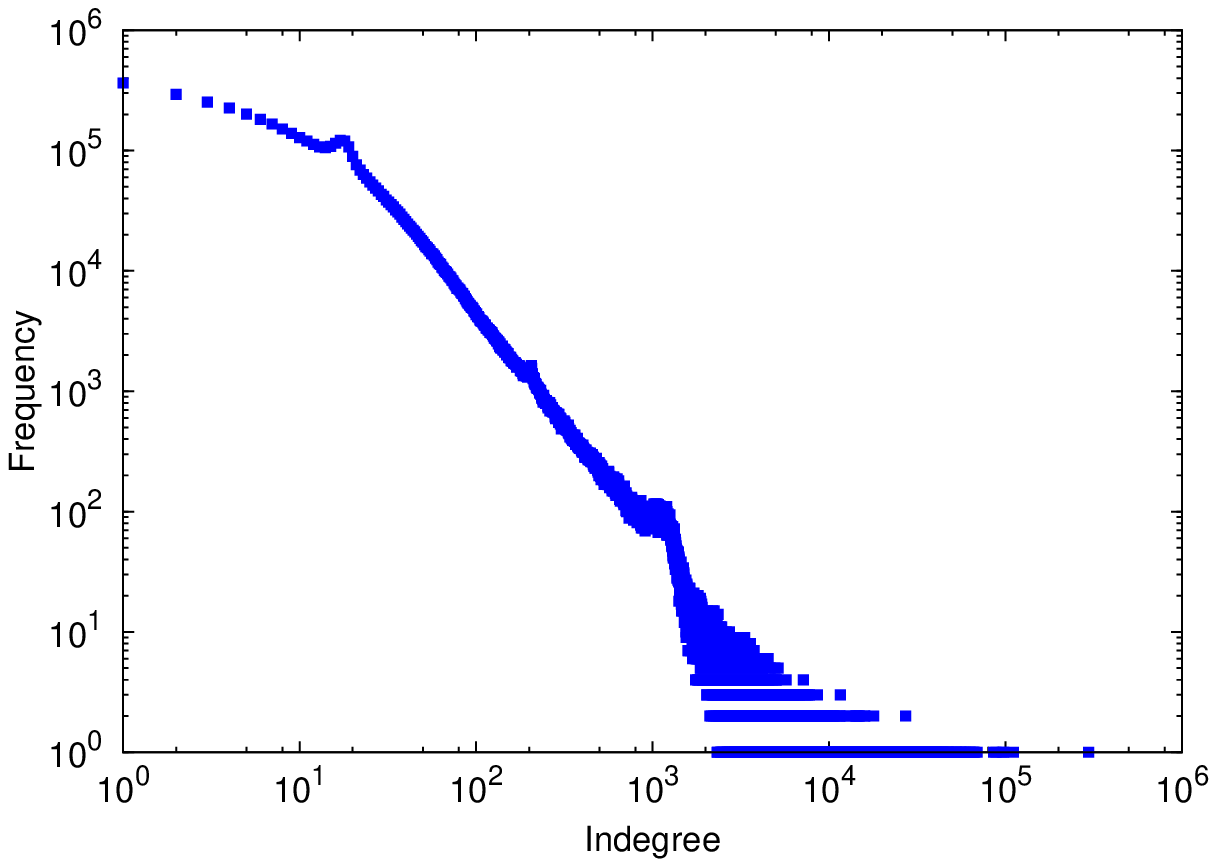,width=5.3cm,height=3.5cm}
\end{minipage} &
\begin{minipage}{5.3cm}
\epsfig{figure=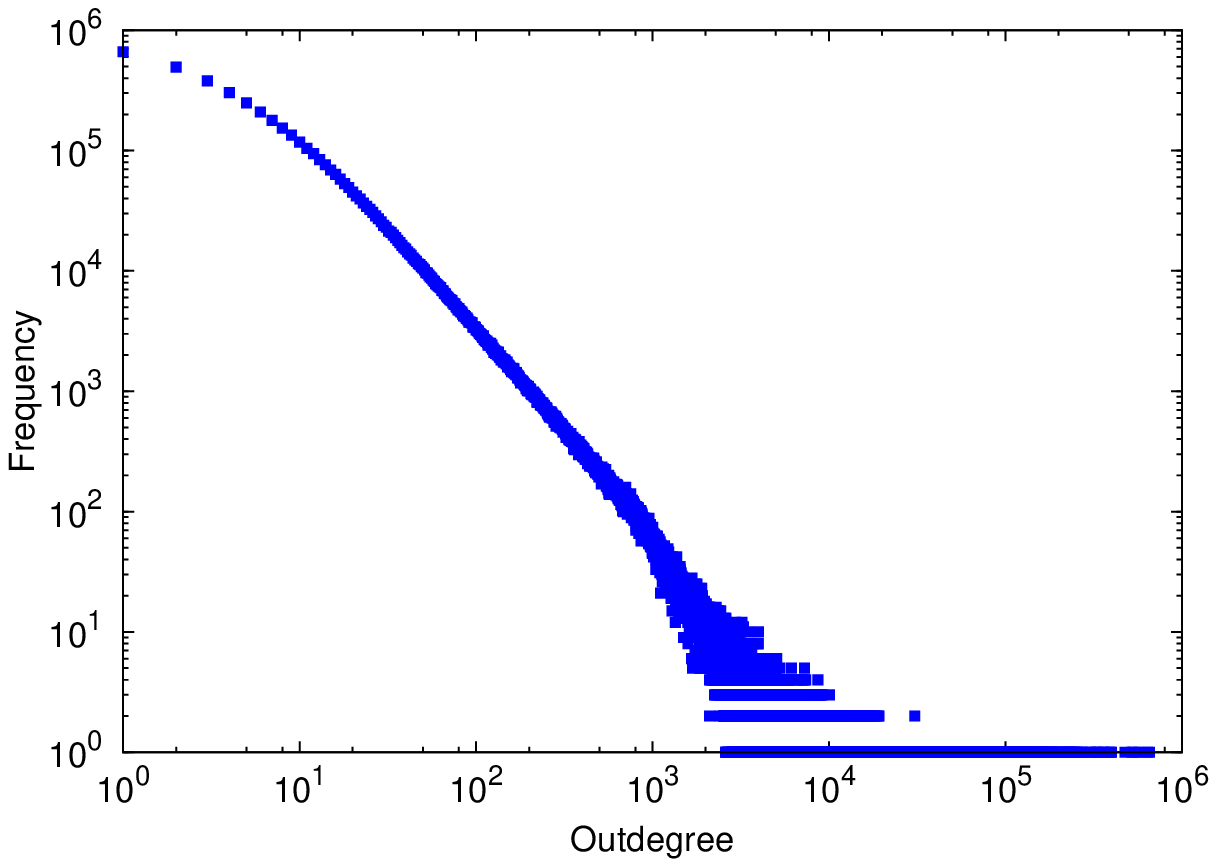,width=5.3cm,height=3.5cm}
\end{minipage} &
\begin{minipage}{5.3cm}
\epsfig{figure=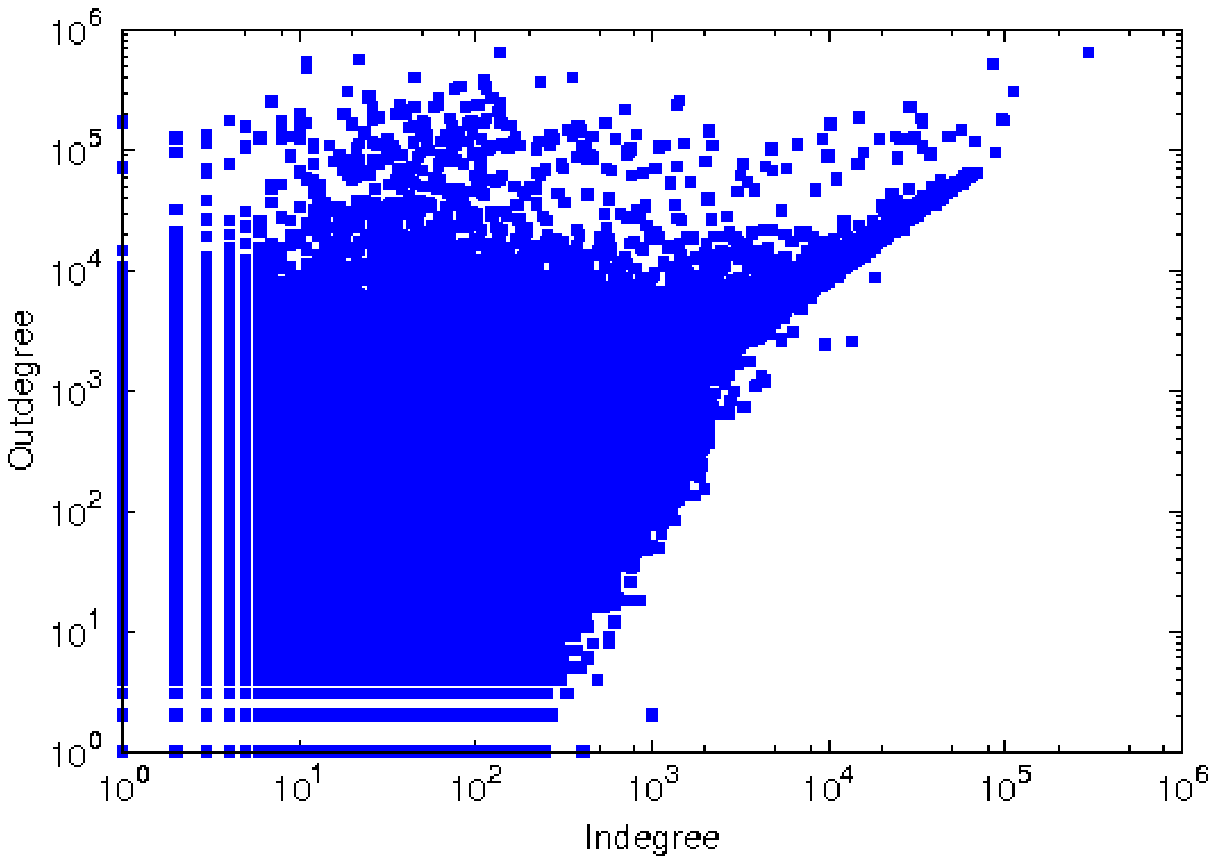,width=5.3cm,height=3.5cm}
\end{minipage} \\
(a) In-degree distribution &(b) Out-degree distribution &(c) In-degree VS Out-degree\\
 \end{tabular}
\caption{Basic properties of the Twitter social network.}
\label{fig:properties}
\vspace{-.1in}
\end{figure*}

Figure~\ref{fig:properties} shows the basic properties of the collected Twitter graph. 
Both the in- and out-degree  distributions follow a power-law~\cite{Kwak_2010}.  Most users follow a few people and a few users 
are followed by a large number of people (``celebrities'').
We notice that the in- and out-degree values 
are positively correlated, as shown in Figure~\ref{fig:properties}(c) by an overall clustering of points around the $x=y$ line.
In addition, we observe an additional cluster of points on the top-left quadrant of this graph, 
which are ``celebrities''. 



\subsection{Geospatial Data}
\label{sec:geospatialdata}

We augmented the tweet7 data set by querying Twitter to obtain user
profile information, which often contains the user's location. Using
the Twitter User ID available in the tweet7 data set, we queried
Twitter using its API to extract location information from user
profiles. The major issue we faced was that of rate limiting by
Twitter. Using a white-listed Twitter account allowed 20,000 queries
per hour, and running parallel instances of our crawler to reach up to
this limit, we were able to retrieve location information of
approximately 7.4 million Twitter users in approximately 200 hours. This is
roughly 75\% of the user population available in the tweet7 data
set. The rest of the users had been banned by Twitter or had deleted
their accounts in the time elapsed between our data collection June, 2011
 and the time the tweet7 data was generated June, 2009.

Only 62\% of the 7.4 million users in the data set provided location
information. A further problem was that location was specified in
several different formats. Hence we first converted the location
information into latitude-longitude pairs and then reverse-geocoded
the coordinates into a city, stat,e and country format, using the
Yahoo! PlaceFinder service which is part of the Yahoo Geo
API.

After processing all available location information, countries with the largest number of users 
in the  data set are the USA (57.6\% users), followed by UK (7.7\%), Brazil (7.1\%), and 
Canada (3.7\%). As the USA contains large number of users, we sub-divide it into 
five commonly used regions: Northeast (10.7\%),  Southeast (13.5\%),  Midwest (10.4\%),
Southwest (6.8\%), and West (16\%).

\subsection{Topic Identification}

The Twitter service allows users to identify topics in messages using hashtags, a single word
starting with '\#'. This allow users to follow not only other users but also conversations around specific 
tags. Using hashtags to identify topics in tweets has been used in several past 
contributions~\cite{Romero_2011,Kwak_2010}, but results in sparse data sets, as the majority 
of users do not use hashtags. From Table~\ref{table:dataset}, we can see that 
only about 10\% of the tweets contain a hashtag. 

In order to determine topics for tweets without hashtags, we used the
OpenCalais~\cite{opencalais_url} text analysis engine to extract
entities, fact, and events from tweet texts in English, French, and
Spanish languages. We have, in a sense, ``outsourced'' the semantic
analysis of tweets to the OpenCalais service. This service extracts
entities and tags. We have assumed that each distinct tag or entity is
a separate topic, although, as we describe subsequently, the output
provided by OpenCalais is just a starting point in the process of
settling on a reasonable set of topics.

OpenCalais also has a rate limit of 50,000 requests per day. This made
sending one tweet a time a virtual impossibility. 
 Using a small sample of text data, we computed the topic
coverage value of a bundle of tweets, by comparing the number of
identified topics when bundling, with the number of identified topic
without bundling. We vary the bundle size from 1 tweet to 100KB worth
of tweets. Using this technique, we determine an optimum bundling size
of 40K, which allows a tractable topic extraction time of 2 weeks,
while resulting in a topic coverage value of approximately 94\%.

Using this method, we were able to extract nearly 6.2M topics in 52M tweets. The remaining tweets were discarded 
as no topic could be identified. Reasons for this include short tweet, unsupported language, or no clearly identifiable
topic. Examples of topics identified and their occurrence count are 'IRANELECTION' (341674), 'TWITPOCALYPSE' (1031),
'SONG OF NICK JONAS' (16).  

\begin{figure}
\centering
\begin{tabular}{cc}
\begin{minipage}{4.cm}
\epsfig{figure=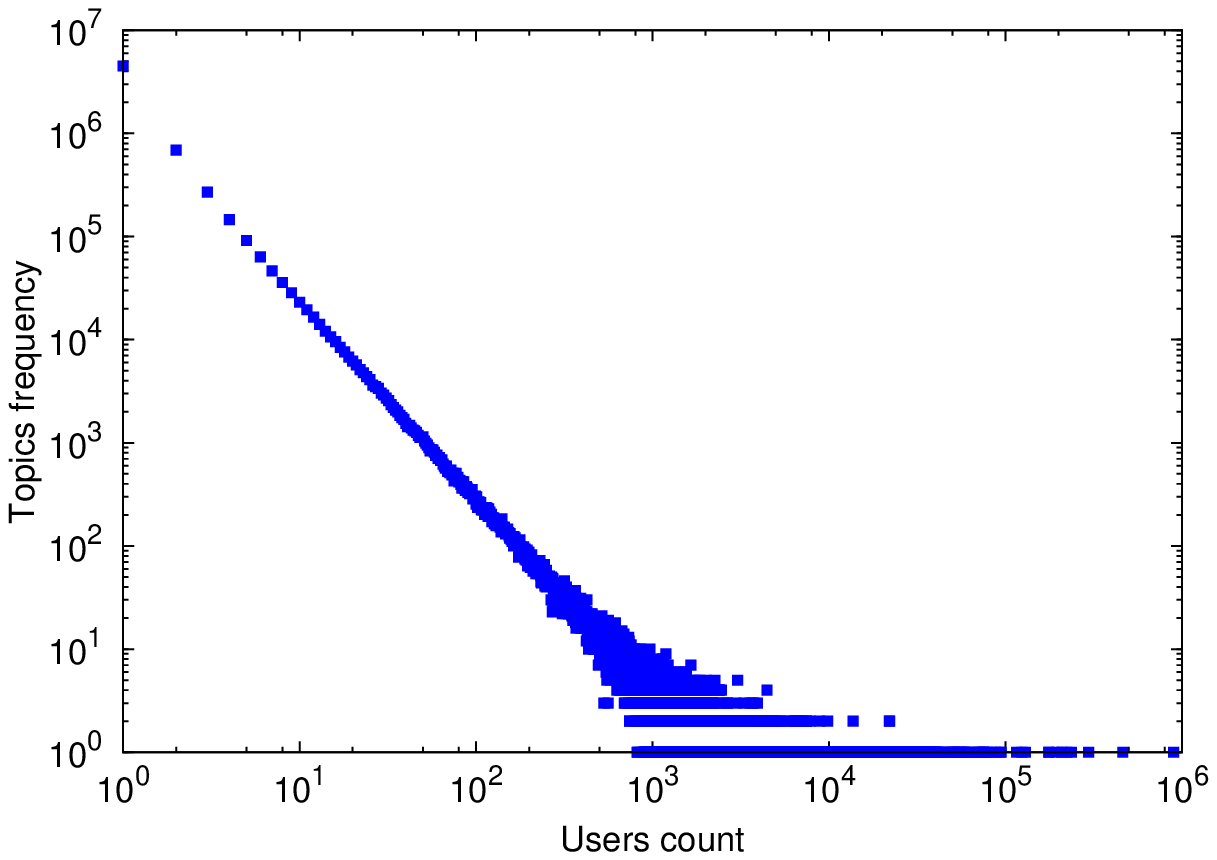,width=4.cm,height=3.5cm}
\end{minipage} &
\begin{minipage}{4.cm}
\epsfig{figure=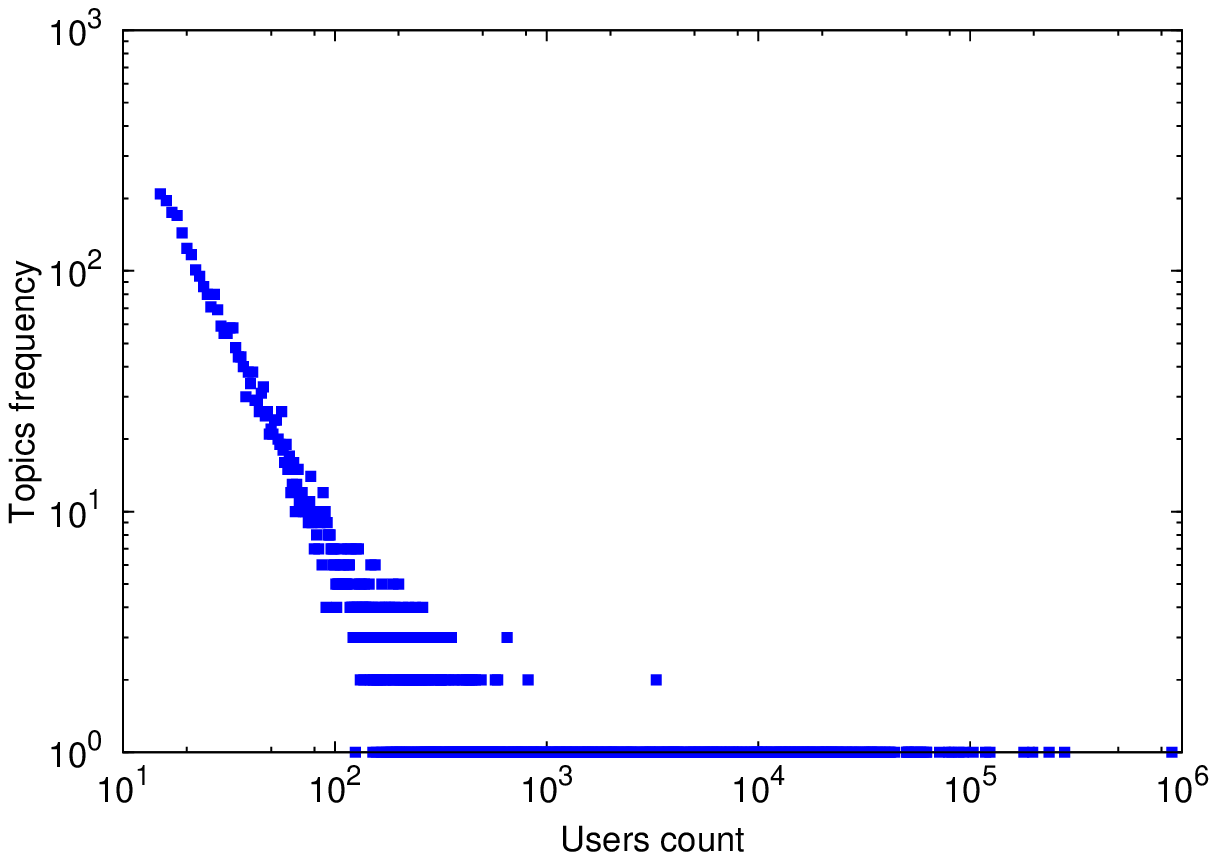,width=4.cm,height=3.5cm}
\end{minipage} \\
(a) Actual topics&(b) Reduced topics\\
 \end{tabular}
\caption{Frequency distribution of topics.}
\label{fig:topics_freq}
\vspace{-.1in}
\end{figure}

Figure~\ref{fig:topics_freq}(a) shows the frequency distribution of
these 6.2M topics.  The distribution follows a power-law shape: most topics are talked by very few users (< 10
users). These topics can be considered as noise in our study, as they
are useless to study the diffusion and popularity of topics in the
network.  We therefore apply a threshold-based filtering, by removing
topics used by less than 15 users. After filtering, we are left with
0.9M topics which is still quite a large set. We further reduce this
set by sampling 4135 topics which were manually examined to ensure
that duplicates were merged and that the topic set comprised sensible
concepts, taking care to include both popular and non-popular topics.
Figure~\ref{fig:topics_freq}(b) shows the frequency distribution of
this reduced topic set. We can see in Figure~\ref{fig:topics_freq}(b)
that, the frequency distribution follows a power-law.  In the rest of
the paper we refer to this set of 4135 as the {\em base set} of
  topics. The analysis done in Sections~\ref{sec:topology}
and~\ref{sec:geography} is done using this set of topics.

\begin{figure}[htbp]
\centering
\includegraphics[width=6cm,height=4cm]{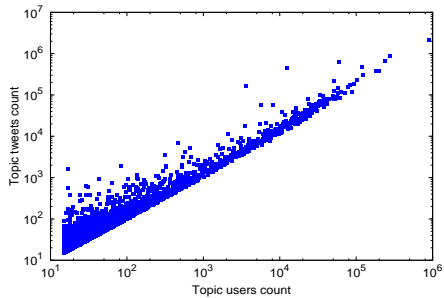}
\caption{Topic popularity diversity.}
\label{fig:users_vs_tweets}
\end{figure}

To identify and measure the popularity diversity of topics, Figure~\ref{fig:users_vs_tweets}
compare the number of users to the number of tweets, by plotting those two variables
on a scatter plot, for each of the 4135 topics. This graph effectively shows the difference
in popularity for all topics. From the graph, we can see that users of unpopular topics typically
tweets more than one time on that topic. Popular topics on the other hand, are typically tweeted 
once by most users. In the following section, we observe and explain why some 
topics become popular, and other not. 
 
One problem with the base set is that it contains topics that already
existed in the network at the beginning of the time window over which
the data set was collected. While this still allows us to observe the
temporal and spatial variation of the topics during the window, it
makes it difficult for us to estimate the effect of initiators on the
popularity of a topic. In order to address this problem filtered out
all topics whose first tweet appears within the first 7 days of the
window. We called this reduced set of topics the {\em filtered
  set}. The analysis of the effect of initiators
(in Section~\ref{sec:initiators}) is done using only the topics in the
filtered set. 

We note that an initial report on the data engineering process
described in this section has been accepted for
presentation~\cite{Ruhela_2011}. However, there have been significant
extensions done for the purposes of this work, including the querying
of follower-following relationships of several million users whose
links were missing from the original data set, and the creation of the
filtered set for the purposes of studying initiator influence.

\section{The influence of initiators}
\label{sec:initiators}

The sudden rise in importance of Twitter as a global communication
medium has made it important to study who are the entities that wield
most influence on this medium. In this section we make an initial
contribution by finding, as stated in Observation
1, that popular topics are generally initiated by users with very high
follower counts. These users are usually either traditional or
web-based news media outlets or media personalities (pop stars,
politicians, writers etc.); we will refer to such
users as {\em celebrities.}  Finding that the mean number of followers
of a user in our data set's Twitter network was 65.7 and the standard
deviation of this quantity was 1291.7, we decided to designate any
user with more than 3,000 followers as a celebrity. 

As stated earlier, we conducted our study on the influence of
initiators on popularity only on the filtered set of topics (i.e., those
whose first tweet appeared at least 7 days after the beginning of our
data set's time window). Popularity here refers to the number of users
tweeting about a topic on a day. For each such topic we considered the first
5\% of all tweets on that topic in the window and designated the set
of users who posted these tweets as the {\em initiators} of the topic.

\begin{figure}
\centering
\begin{tabular}{cc}
\begin{minipage}{4.cm}
\epsfig{figure=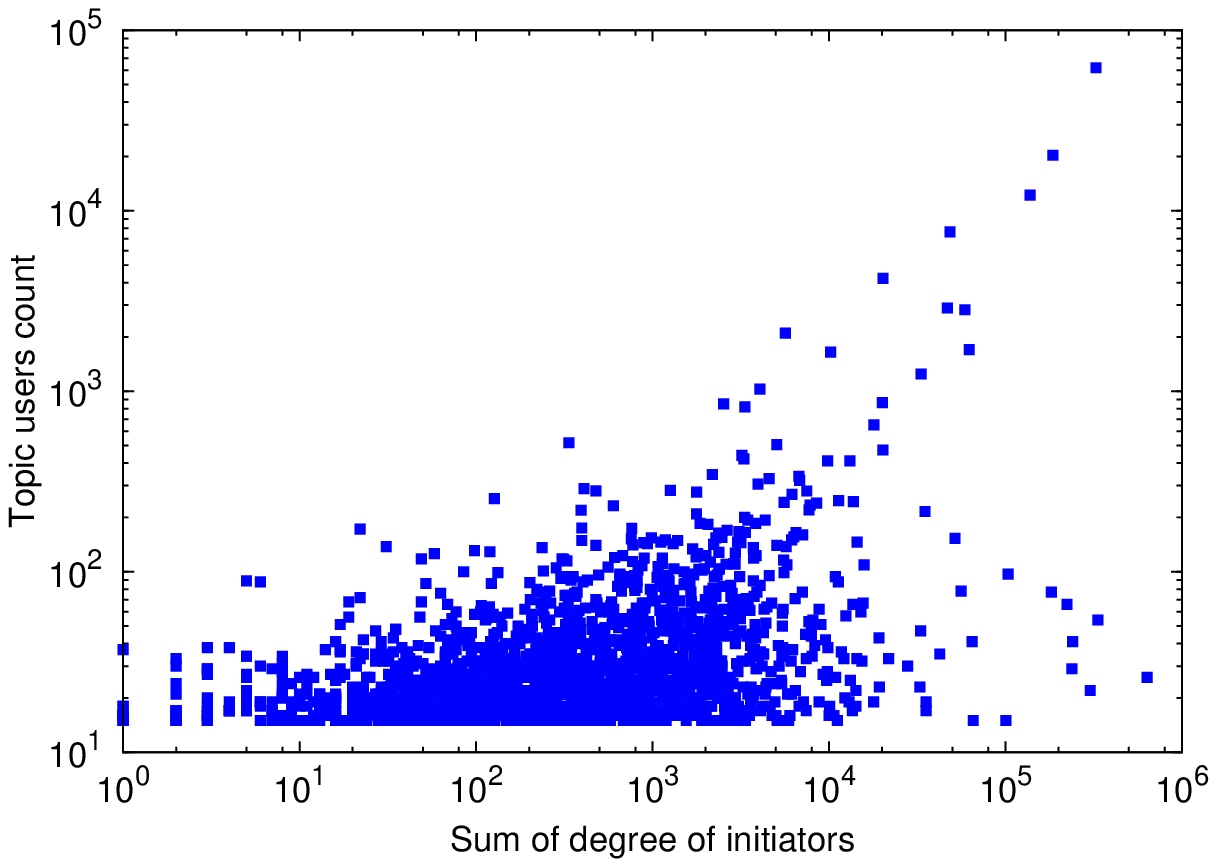,width=4cm,height=3.4cm}
\end{minipage} &
\begin{minipage}{4.cm}2
\epsfig{figure=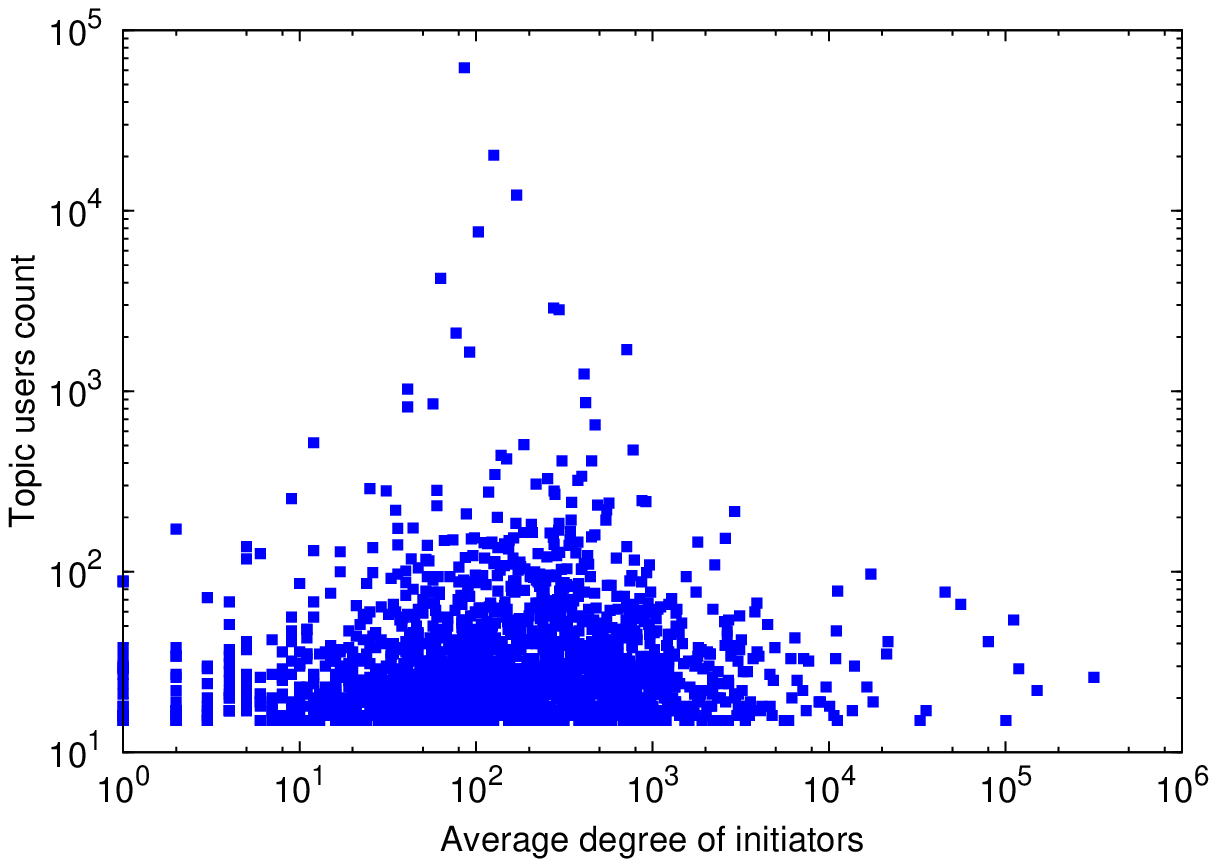,width=4cm,height=3.4cm}
\end{minipage} \\
(a) Total \#followers & (b) Average \#followers \\
 \end{tabular}
\caption{\#followers of initiators versus popularity I.}
\label{fig:initiator_degree}
\vspace{-.1in}
\end{figure}

In Figure~\ref{fig:initiator_degree} we present two scatter plots: the
first shows the relationship between popularity and the total number of
followers of the initiators. We note that highly popular topics have
very high aggregate followers of the initiators. But we note in
Figure~\ref{fig:initiator_degree}(b) that the average number of followers of the
initiators of highly popular topics is in the hundreds rather than the
thousands. This indicates that there are some initiators with very large number of  
followers involved in these popular topics. 

\begin{figure}
\centering
\begin{tabular}{cc}
\begin{minipage}{4.cm}
\epsfig{figure=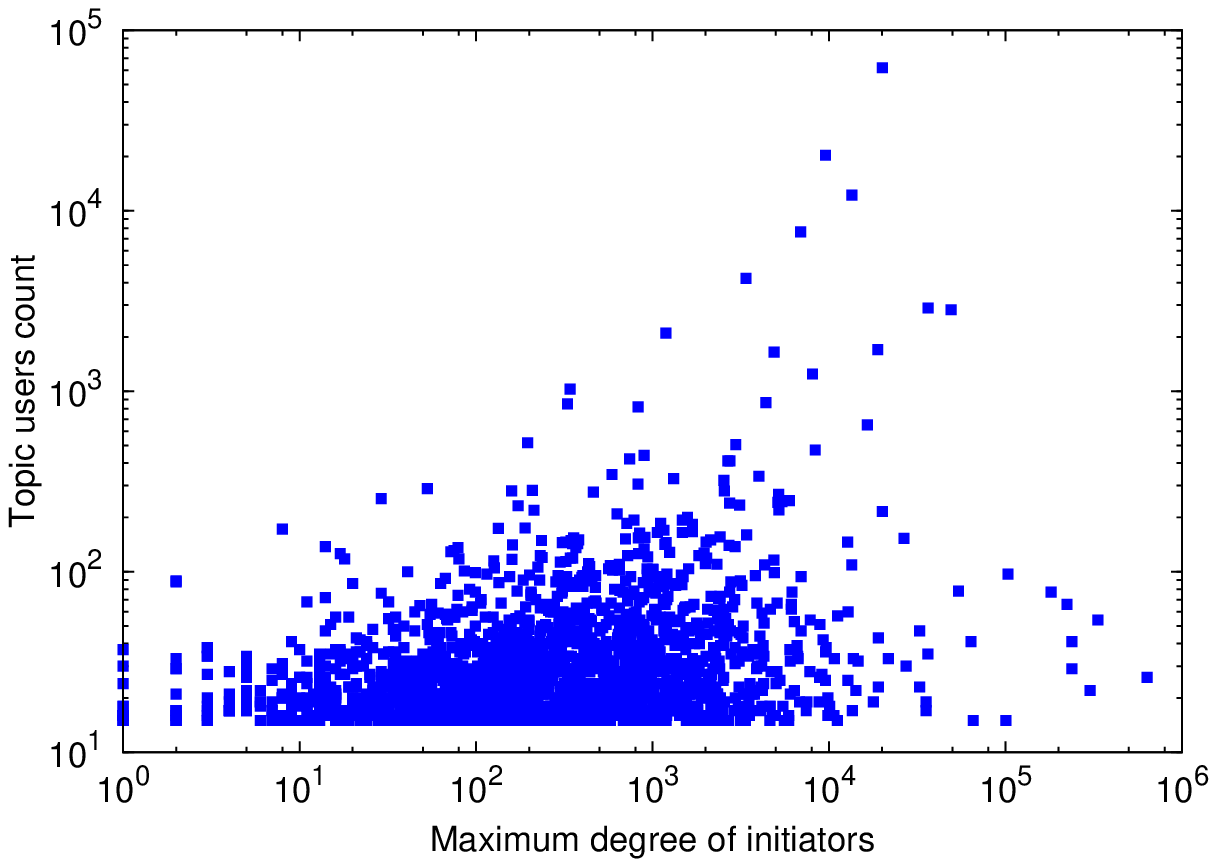,width=4.cm,height=3.4cm}
\end{minipage} &
\begin{minipage}{4.cm}
\epsfig{figure=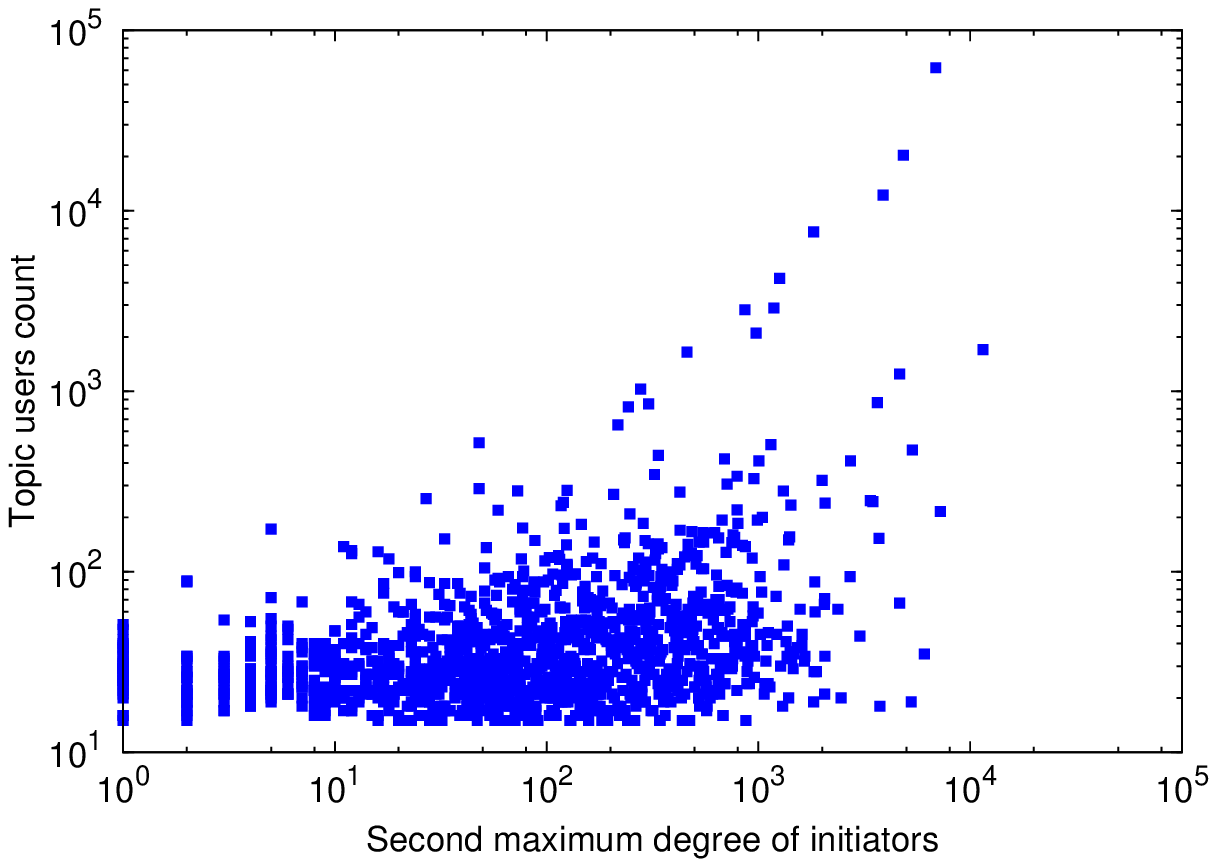,width=4.cm,height=3.4cm}
\end{minipage} \\
(a) Maximum \#followers  &(b) 2nd highest \#followers \\
 \end{tabular}
\caption{\#followers of initiators versus popularity II.}
\label{fig:initiator_degree_max}
\vspace{-.1in}
\end{figure}
When we examined the maximum and second highest number of followers of
the initiators of each topic (cf. Figure~\ref{fig:initiator_degree_max}) we
found that it was indeed the case that celebrity users were involved
in initiating highly popular topics while most unpopular topics were
initiated by users with a low number of followers. 

An interesting observation can be made by looking at the points
plotted near the bottom right corner of all plots in
Figures~\ref{fig:initiator_degree}
and~\ref{fig:initiator_degree_max}. These are topics started by a few
celebrities that did not achieve any popularity. Hence we see that
while it is the case that celebrities drive the popularity of topics,
it is not the case that every topic promoted by celebrities becomes
popular.  This helps us establish Hypothesis 1: Celebrities influence
the spread of topics, but cannot make a topic popular unless common
users pick up on them.

\begin{figure}[!t]
\centering
\includegraphics[width=6cm,height=4cm]{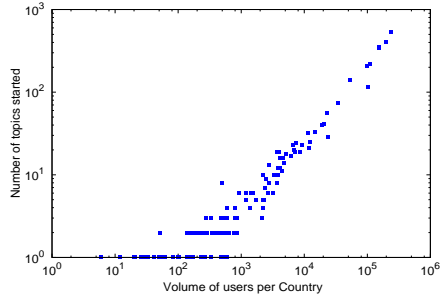}
\caption{Users vs topics initiated per country.}
\label{fig:UsersVsTopicsIni}
\end{figure}

We expect that regions containing larger numbers of Twitter users will
influence the topics discussed to a greater extent. To determine this
we tabulated the number of topics for which each region has at least
one user in the initiator set (cf. Figure~\ref{fig:UsersVsTopicsIni}).
We find that there is a linear relationship between the number of
topics initiated and the size of the regions user base. While this
does lead us to conclude that regions with larger user bases have more
influence on the network, at least the relationship is not
super-linear, implying that an increase in users in a given region
could potentially lead to an increase in the share of the topics
initiated therein.

\begin{figure}[!t]
\centering
\includegraphics[width=6cm,height=4cm]{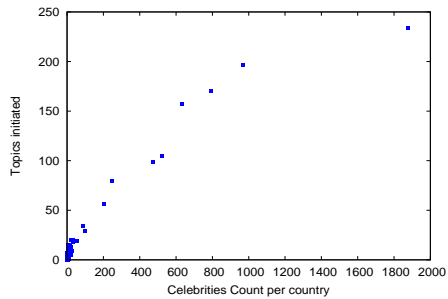}
\caption{Topics started  vs Celebrities count per country.}
\label{fig:CelebsVsTopicsIni}
\end{figure}
However, a cautionary note is struck in Figure~\ref{fig:CelebsVsTopicsIni}.
Here we again plotted the number of topics started by a region on the
$y$-axis, but on the $x$-axis instead of plotting the size of the user
group from the region (as done in Figure~\ref{fig:UsersVsTopicsIni}),
plotted the number of celebrities contained in those regions. Further,
we did not consider all topics for this plot, instead focusing on
only the top 500 topics (by topic user count)  in the
filtered set. We found that countries with a greater number of
celebrities initiated a disproportionately high number of these
topics.

We see a kind of continuity with the past here. The cultural and
political dominance of certain regions that existed before Twitter
came into being is reflected in the presence of a greater number of
celebrity users in those regions, and consequently translates into a
greater impact for those regions in terms of popular topics.

\section{Graph theoretic properties of topics} 
\label{sec:topology}

In this section we establish Hypothesis 2 and argue towards
Hypothesis 2a. To do this we study three different types of graphs
associated with each topic. First, we study the {\em
  lifetime graph} of a topic; this is the subgraph induced on the Twitter
network by all the users who have tweeted on that topic at any time in
our window. Second, we study the {\em evolving graphs} of a topic. In particular, we
partition the tweets related to a particular topic by day and for each
day we construct the subgraph induced on Twitter by the users who have
tweeted on that topic on that particular day. Finally, we study the
{\em cumulative evolving graphs} of a topic. We denote by $G_i^t =
(V_i^t,E_i^t)$, the cumulative evolving graph for topic $t$ on day $i$
and define it as follows:
\begin{itemize}
\item  The vertex set of $G^t_0$  
comprises the users $V_0$ who tweet about $t$ on day 0. The edge set
is empty.
\item The vertex set $v_i^t$ of $G_i^t$ is the set of all users who have
  tweeted on a topic in days 0 through $i$. An edge $(u \rightarrow
  v)$ is added to $E_i^t$ if $u \in V_{i-1}^t$ and $v$ tweets about
  $t$ on day $i$. 
\end{itemize}

\subsection{Lifetime Graphs}
\label{sec:topology:lifetime} 

\begin{figure}[tbp]
\centering
\includegraphics[width=6cm,height=4cm]{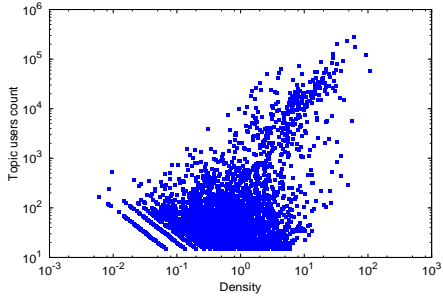}
\caption{Lifetime graphs: Density vs popularity.}
\label{fig:topic_graph}
\end{figure}

We constructed lifetime graphs for each of the 4135 topics in the base
set. Our first observation from these graphs is that popular topics
tend to occupy the more well-connected portions of the network. To
establish this we studied the relationship of the total number of
users who have tweeted on a topic (referred to as the {\em topic user
  count}) to the {\em density} of the lifetime graph of the topic (defined as 
the number of edges per user in the graph). 
In Figure~\ref{fig:topic_graph}, we note something important: the
lifetime graphs of non-popular topics (e.g., user count < 1000) do not have densities greater
than 10, and in fact many tend to have a density less than
1. A density of less than one for a subgraph of a reasonably
well-connected graph like the Twitter network clearly indicates a high
number of small isolated clusters. This isolation is observed even
in the lifetime graph which establishes relationships between users
even where they may not exist, for example, by putting an edge from $u$ to
$v$ although $v$ may have tweeted on the topic {\em before} $u$. Hence
Figure~\ref{fig:topic_graph} strongly supports one side of Hypothesis
2: less popular topics generally exist in highly disconnected
clusters.

\begin{figure}[thb]
\centering
\includegraphics[width=6cm,height=4cm]{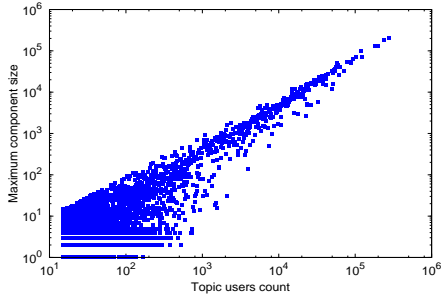}
\caption{Lifetime graphs: Largest connected component vs popularity.}
\label{fig:giant_comp}
\end{figure}

It is difficult to establish the other direction of Hypothesis 2 from
lifetime graphs because of the optimistic selection of edges mentioned
earlier. Nonetheless we get strong indicative evidence for our
hypothesis that a popular topic tends to be discussed in one large
cluster that contains most of the users that have tweeted on that
topic. This evidence comes from studying the relationship of the topic
user count to the size of the largest connected component of the
lifetime graph, as shown in Figure~\ref{fig:giant_comp}. From  Figure~\ref{fig:giant_comp},
notice that for more popular topics there is a clear linear relationship
between the popularity of the topic and the size of the largest
component of the lifetime graph. This is strongly indicative of
Hypothesis 2, although it cannot be used as conclusive evidence.

\begin{figure}[htbp]
\centering
\includegraphics[width=6cm,height=4cm]{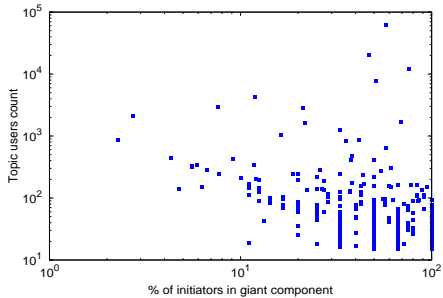}
\caption{Initiators in giant component.}
\label{fig:initiator_component}
\end{figure}

As an aside, we investigated how many of the initiator nodes are
present in the largest connected component for the topics in the
filtered set used in Section~\ref{sec:initiators}. 
In Figure~\ref{fig:initiator_component}, we see that the more popular
topics tend to contain almost all their initiators in their largest
connected component (which we know from above is very large). Hence
popular topics tend to spread around their initiators, implying that
initiators have a topological impact as well as a geographical one for
popular topics. Another way of putting this is that those people who
are not connected to the ``cool'' people may in fact miss out on the
``hot'' topics: an unfortunate conclusion that leads us to think that
greater awareness gives some users a jumpstart in the process of
collecting cultural capital on OSNs like Twitter.

\subsection{Evolving Graphs}
\label{sec:topology:evolving}

It is in the study of evolving graphs that we are able to establish
that most users tweeting on popular topics form one large connected
component (we will refer to this large component as the {\em giant
  component} from now on). To establish this we began by splitting our
base set into three categories, depending on the topic user count
$\rho$: \emph{popular} ($\rho>10,000$), \emph{medium popular}
($1000< \rho <10,000$) and \emph{non-popular} ($\rho<1000$). For each
category, we randomly chose $40$ topics, and computed evolving graphs
for each. For each day's graph, we then computed the ratio between the
sizes of the largest and the second largest component, and also the
ratio between the radii of the largest and second largest component.

\begin{figure}[htbp]
\centering
\subfigure[Ratio of components size]{
{\includegraphics[width=6cm,height=4cm]{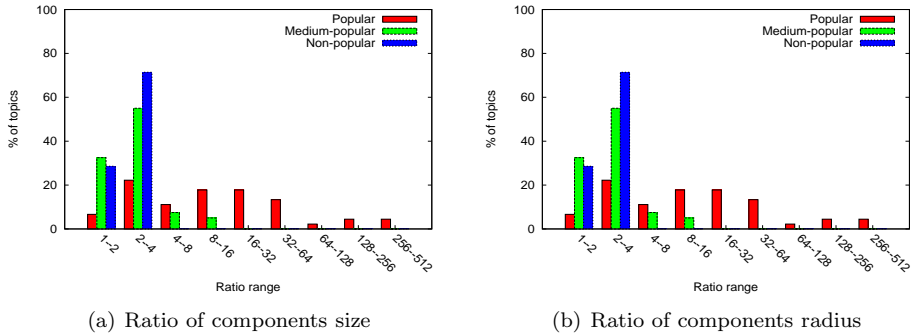}}
}
\subfigure[Ratio of components radius]{
{\includegraphics[width=6cm,height=4cm]{figures/median_comp.eps}}
}
\caption{Median value of the ratios.}
\label{fig:ratio-median}
\vspace{-.1in}
\end{figure}

In Figure~\ref{fig:ratio-median}(a) and (b) we present histograms
for the ratios of component and radii sizes. The buckets divide the ranges of ratios
observed for the size and the radius. We find the median ratio for
each topic and display the percentages of these medians that land in
each bucket. Note that only the highly popular topics populate the
buckets with size ratio greater than sixteen, and that the median size
ratio for these popular topics goes all the way up into the range of
$10^2$ and this is just the median, the maximum tends to be much higher
but we study the median here because it is a more robust
statistic. Most unpopular topics stay below 4 showing a remarkable
evenness in the distribution of component sizes. The radii ratios
similarly show that the width of the reach of the popular topics comes
from the width of one large component rather than from a large number
of small components. This effectively establishes Hypothesis 2.

Moving towards Hypothesis 2a, we first clarify in the context of
evolving graphs what we mean when we say clusters merge. If we
visualize the social network as a set of communities connected through
users who may belong to multiple communities, our narrative of topic
spread says that topics that are going to become very popular witness
intense discussion {\em within} communities at first. When the level
of intensity rises then the users who bridge communities enter the
discussion in a big way causing a merging of what were earlier
disjoint discussions. If Hypothesis 2a is correct then it can be
reinterpreted to mean that the bridge users serve as a baraometer of
the topics rising popularity.  To investigate the applicability of
this narrative we study the conductance of evolving topic
graphs. Motivated by a definition widely used in the study of mixing
times of random walks in graphs, we define the conductance $\phi(S)$
of a subset of nodes $S$ of a directed graph $G=(V,E)$ as the ratio of
the edges outgoing from the vertices of $S$ that land outside $S$:
\[ \phi(S) = \frac{|\{(u \rightarrow v) : u\in S, v \in V\setminus
    S\}|}{|\{(u \rightarrow v) : u\in S\}|}.\]
Clearly, the higher the value of $\phi(S)$, the more the number of
nodes outside $S$ that are made aware of a topic being tweeted by the
users in the set $S$.
\begin{figure*}
\centering
\subfigure[CAMBRIDGE] {
{\includegraphics[width=7cm,height=4cm]{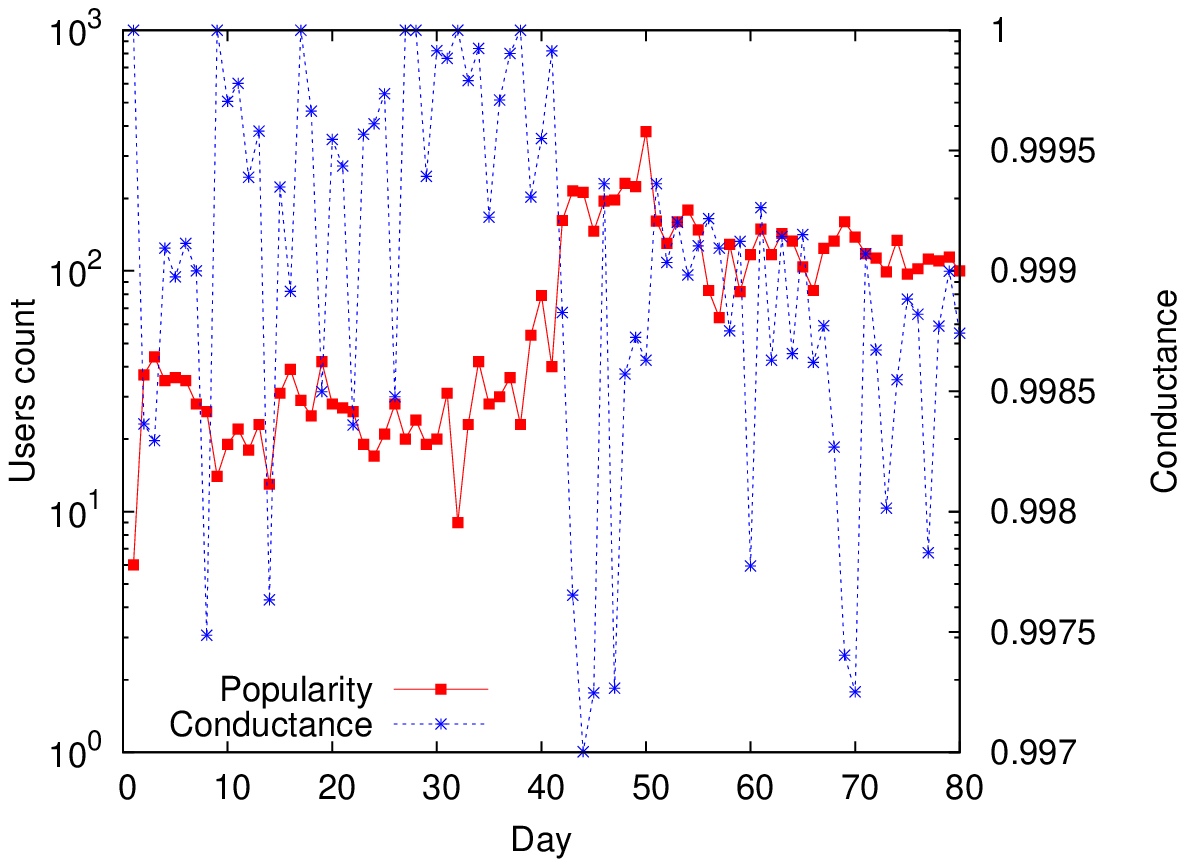}}
}
\subfigure[FOLLOWFRIDAY] {
{\includegraphics[width=7cm,height=4cm]{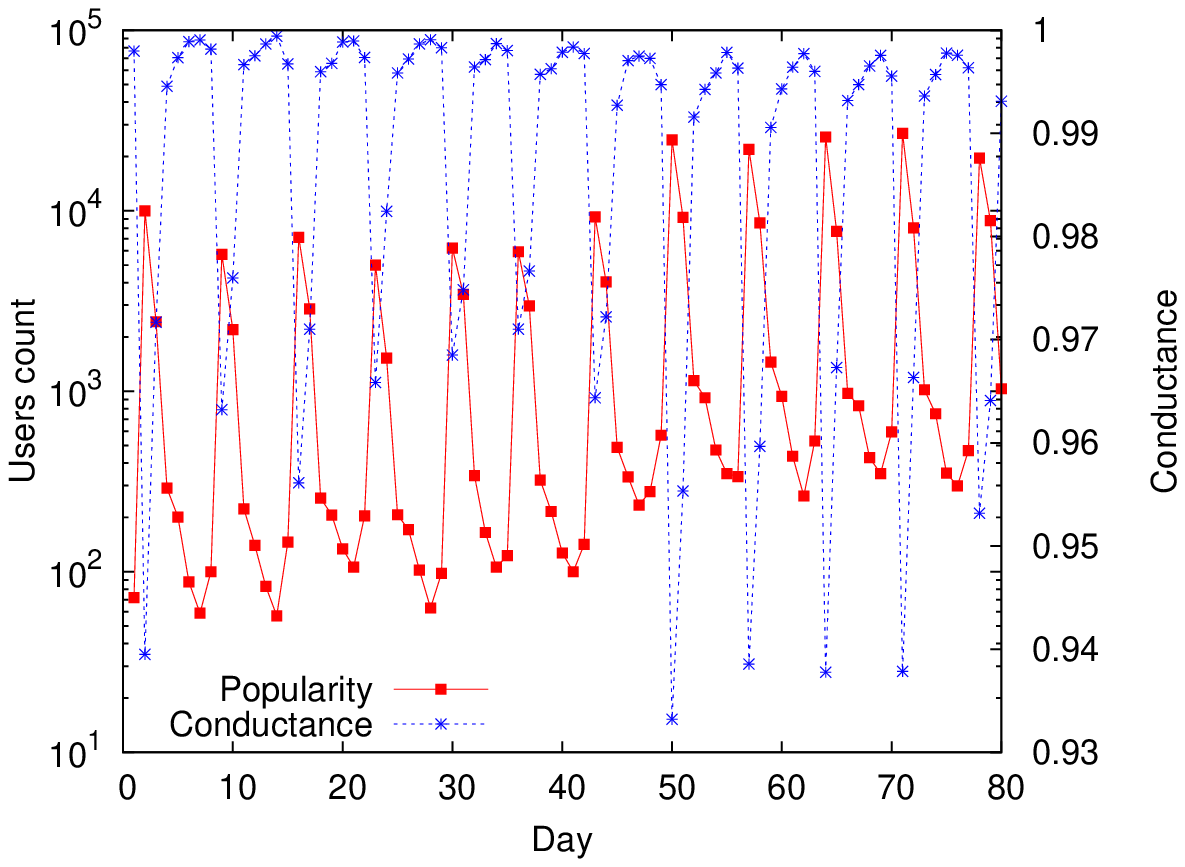}}
}
\subfigure[IRANELECTION] {
{\includegraphics[width=7cm,height=4cm]{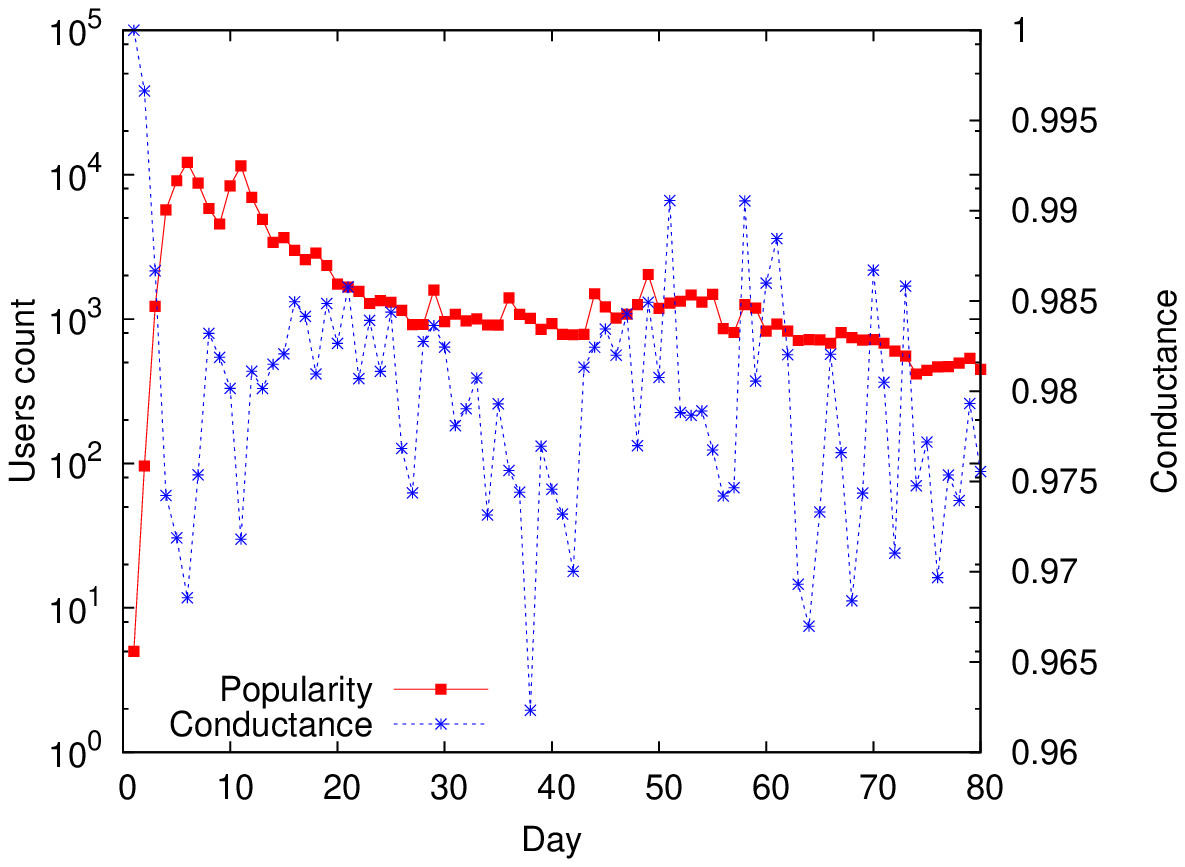}}
}
\subfigure[MICHAEL JACKSON]{
{\includegraphics[width=7cm,height=4cm]{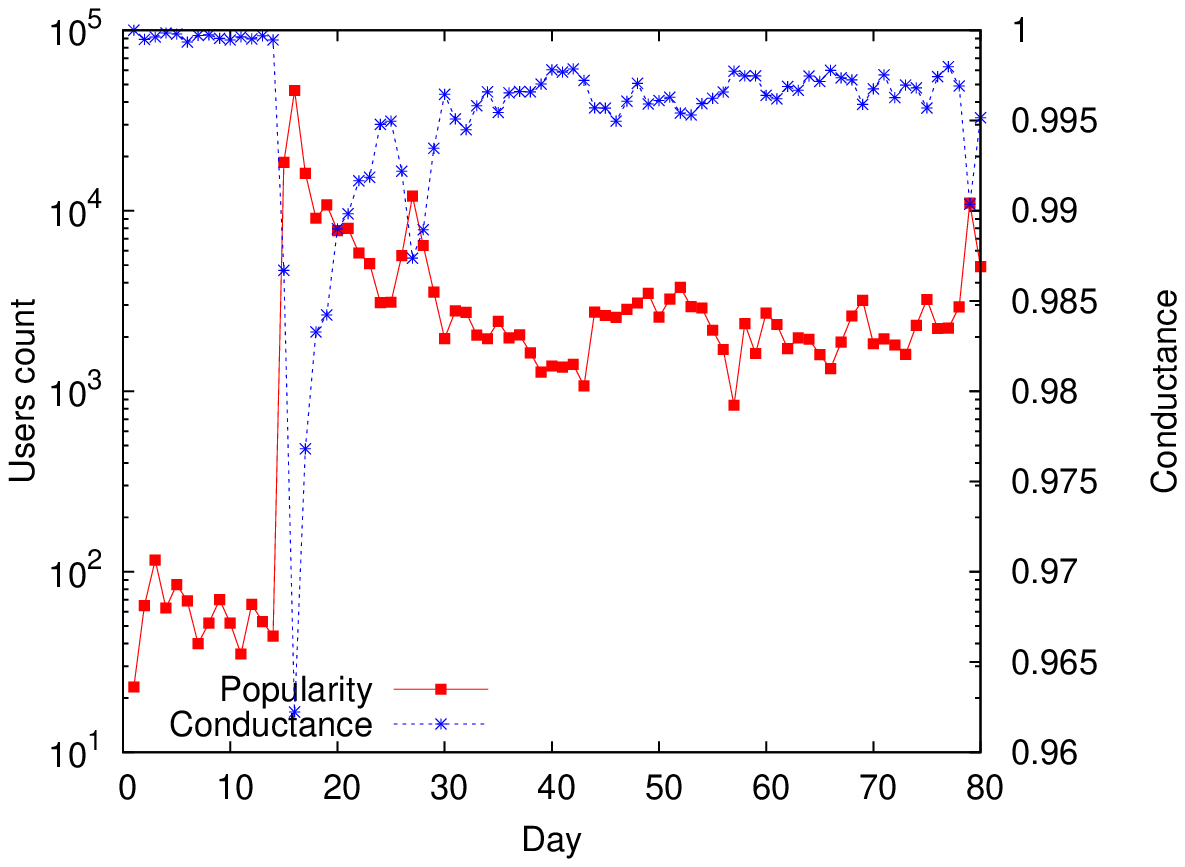}}
}
\caption{Evolving graph conductance.}
\label{fig:conductance_daywise}
\end{figure*}

In Figure~\ref{fig:conductance_daywise} we plot the evolving value of
the conductance of the user set of the day's graph alongside the
evolving topic user count for four topics: one less popular topic
``CAMBRIDGE'', one periodically popular topic ``FOLLOWFRIDAY'', and
two topics that display distinct and very high peaks in their
popularity ``MICHAEL JACKSON'' and ``IRANELECTION''.  Observing the
three popular topics we notice that conductance is very high just
before the peak is seen.  As soon as the peak is formed the
conductance dips down to a low value. This supports Hypothesis 2a
because when the users that bridge distinct clusters start tweeting on
the topic then a larger number of edges become internal to the day's
topic graph, hence the conductance should dip as it does. Again, we
clarify that this result is merely indicative of Hypothesis 2a.

There are a number of other interesting artifacts that can be observed
here. The sharp peak in Figure~\ref{fig:conductance_daywise}(d) comes
on the day of Michael Jackson's demise. The conductance for this topic
was uniformly high earlier, indicating a steady level of discussion
about Michael Jackson, in tune with his general popularity. But his
death leads to a sharp rise in tweets about him, causing an immediate
dip in the conductance. After this initial dip the conductance rises
again but no peak comparable to the first one is see, indicating that
a high sustained level of interest in this topic is accompanied by a
high sustained level of disinterest in the followers of the users
continuing to tweet about Michael
Jackson. Figure~\ref{fig:conductance_daywise}(c) shows a similar
initial behavior accompanying an event, the holding of elections in
Iran. Subsequently there is sustained discussion which is more of the
nature of a conversation (the latter part of the ``IRANELECTION''
trajectory shows an unusually high number of tweets per user on this
topic). This conversation proceeds in regions of the network that have
reasonably high conductance but occasionally show dips in conductance,
indicating a higher level of clustering in the user set, something
that might be expected of a conversation. We note the the similarly
high values of conductance displayed by the topic ``CAMBRIDGE'' in
Figure~\ref{fig:conductance_daywise}(a) have a different connotation
to the high values seen in the other graphs because, like most less
popular topics, this too shows highly disconnected daily graphs.

The results on conductance presented here are replicated in other
evolving graphs for popular topics, but we omit a wider discussion due
to space constraints.

\subsection{Cumulative Evolving Graphs} 
\label{sec:topology:cumulative}

\begin{figure*}
\centering
\subfigure[INDIANA, UNITED STATES] {
{\includegraphics[width=7cm,height=4cm]{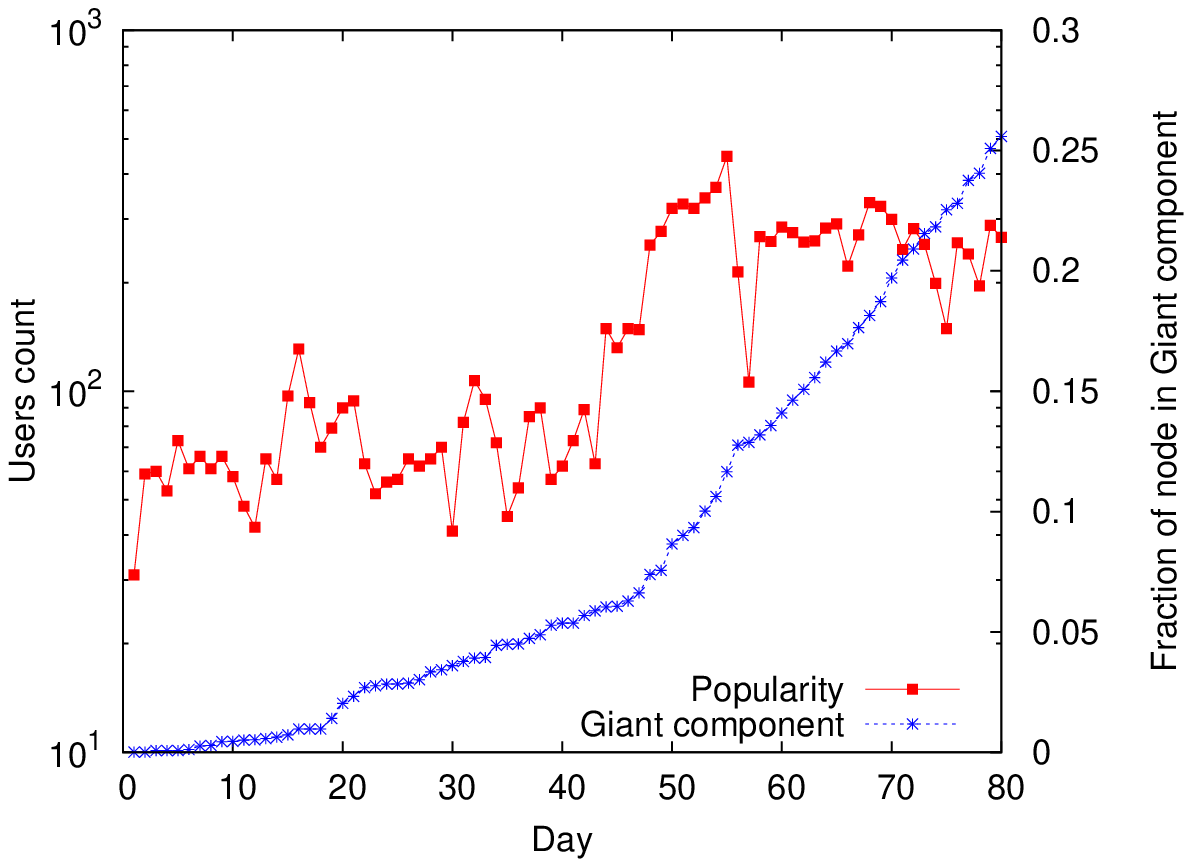}}
}
\subfigure[SMARTPHONE] {
{\includegraphics[width=7cm,height=4cm]{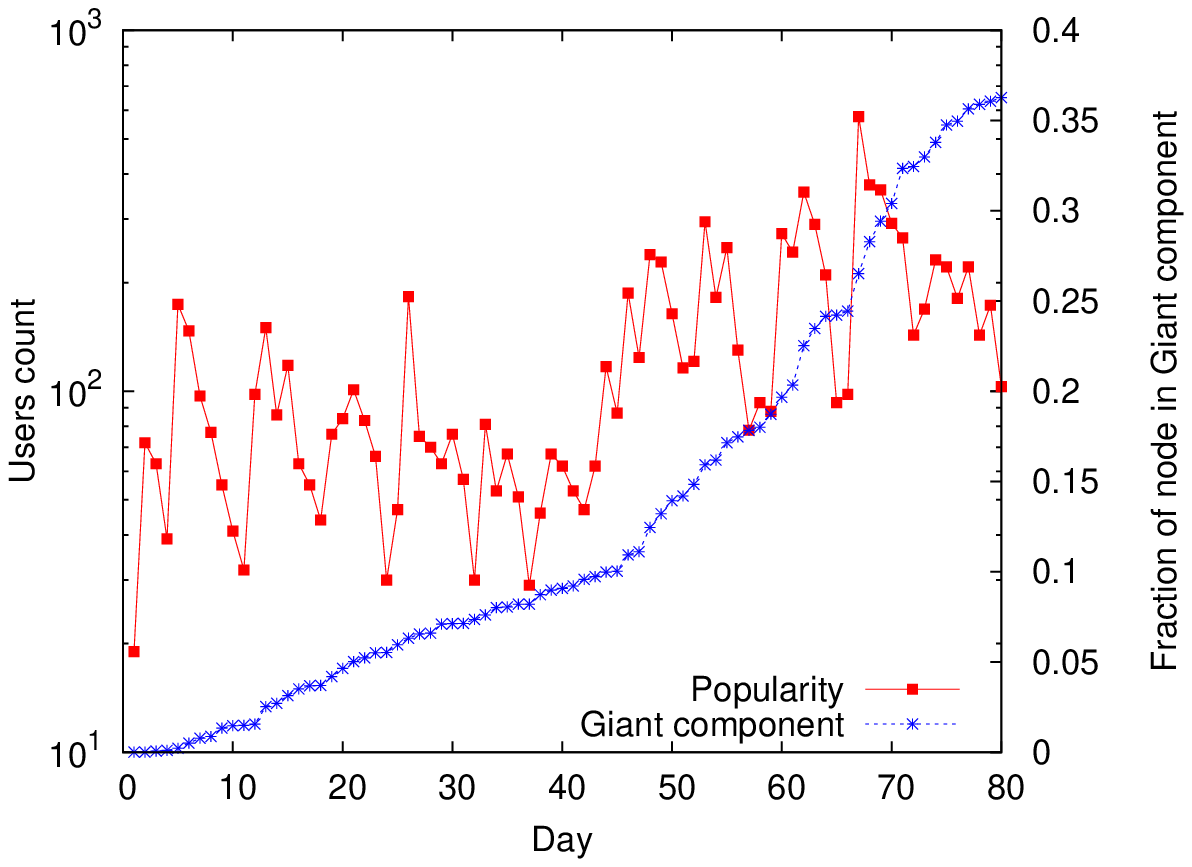}}
}
\subfigure[IRANELECTION] {
{\includegraphics[width=7cm,height=4cm]{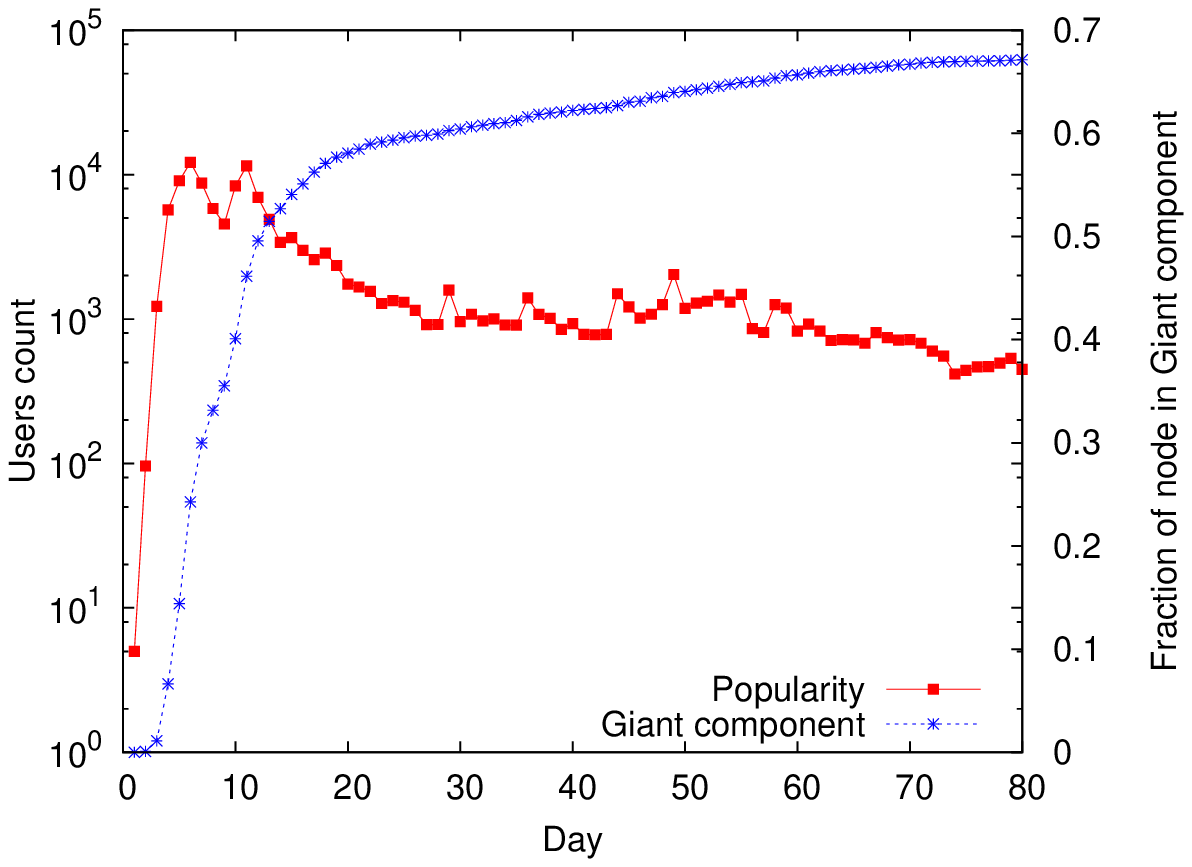}}
}
\subfigure[MICHAEL JACKSON]{
{\includegraphics[width=7cm,height=4cm]{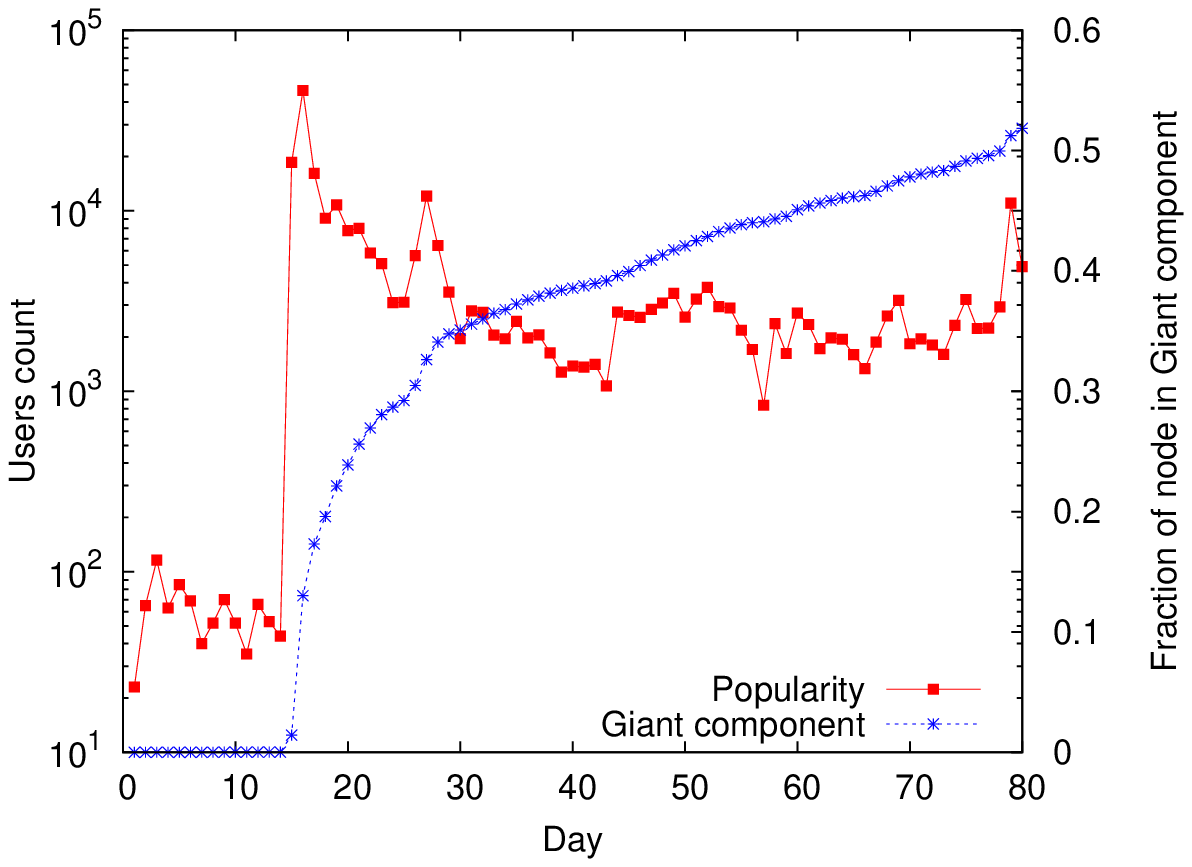}}
}
\caption{Giant component evolution over time.}
\label{fig:time-comp}
\end{figure*}

By using a timing relationship to establish edges for the construction
of cumulative evolving graphs, we make them a better approximation for
the spread of a topic than the lifetime graphs we studied in
Section~\ref{sec:topology:lifetime}. In
Figure~\ref{fig:time-comp} we plot the fraction of nodes
in the largest component of the cumulative evolving graph for two
highly popular topics ``MICHAEL JACKSON'' and ``IRAN ELECTION'' and
two less popular topics ``INDIANA, UNITED STATES'' and
``SMARTPHONE''. Note that even at the end of their evolution the two less popular topics
have only 25\% and 35\% of their users in the giant component of the
cumulative evolving graph while the two popular topics have half the
users in the giant component even before the time window
finishes. This supports Hypothesis 2 since the cumulative evolving
graph is a better approximation of the spread of a topic.

But, more importantly, the sharp rise in the fraction of nodes in the
giant component that accompanies a peak in the evolution of the number
of tweets stands in support of Hypothesis 2a because a merging of
smaller clusters into one large cluster would be accompanied by a
sharp rise in this fraction. It could be argued that this rise in the
fraction is because of a sharp growth in the number of users in the
largest component rather than a merging of clusters, but that seems
unlikely given the extent of the rise, and the large number of users
already present in the cumulative evolving graph at that point. The
less popular topics also show increases in the fraction when their
topic user counts drift upwards, but the rise is much less dramatic
than that shown by the popular topics, and could possibly be explained
by a general growth in the larger component rather than a radical
merging of smaller clusters.

\section{Geographical analysis}
\label{sec:geography}

\begin{figure}[!t]
\centering
\includegraphics[width=6cm,height=4cm]{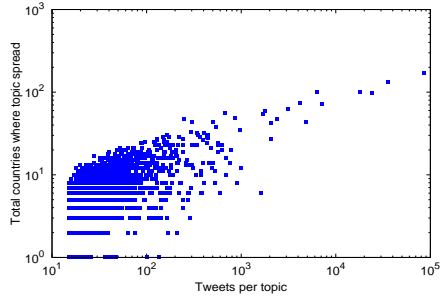}
\caption{Tweets vs regions count.}
\label{fig:TweetsVsCountriesCount}
\end{figure}

This section establishes Hypothesis 3. 
We argue that the popularity of topics is correlated with their
geographical spread. 
We begin by simply studying the number of regions represented by at
least one user talking about a topic and plotted it against the
popularity of the topic (see Figure~\ref{fig:TweetsVsCountriesCount}). It is quite clear from this
plot that the number of regions touched by less popular topics is less
than those touched by more popular topics. This plot does not
establish our hypothesis but it is indicative of it in the sense that
it does not falsify it either.

\begin{figure*}[t]
\centering
\begin{tabular}{ccc}
\begin{minipage}{5.3cm}
\epsfig{figure=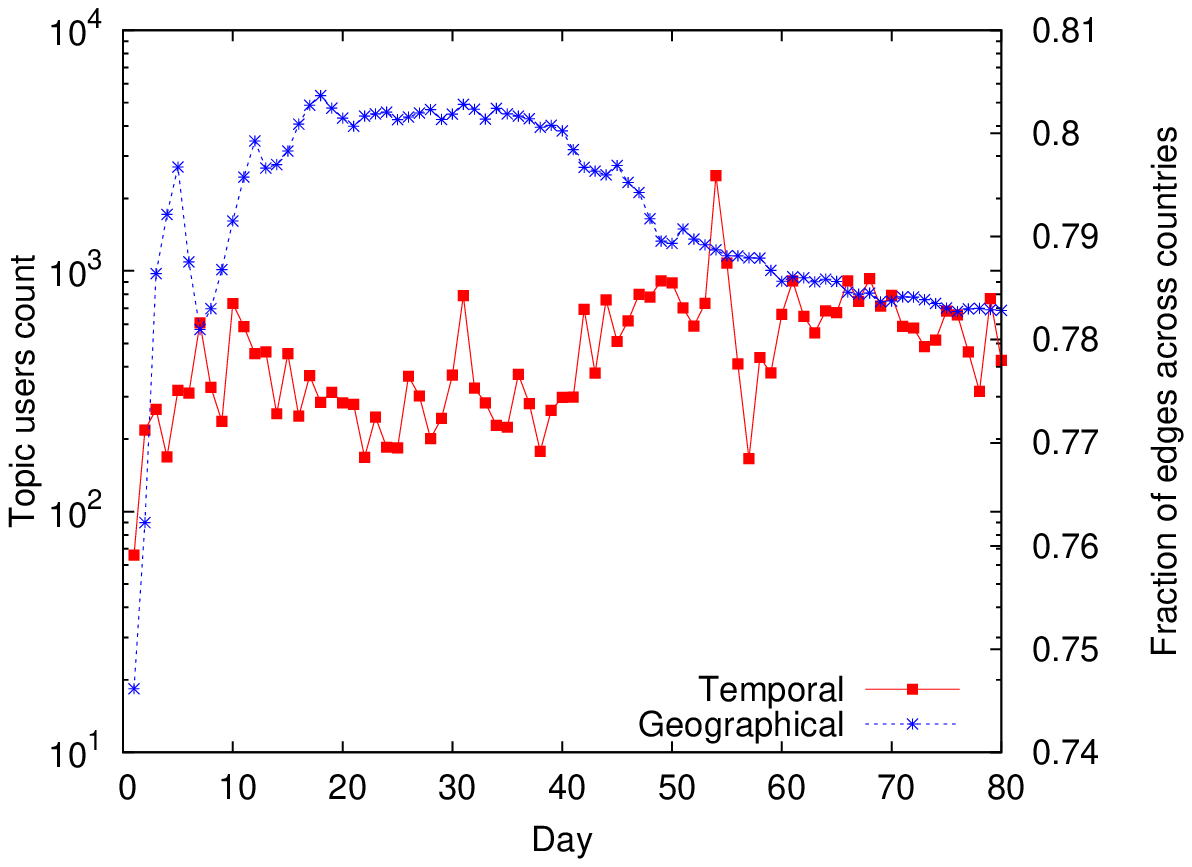,width=5.1cm,height=3.4cm}
\end{minipage} &
\begin{minipage}{5.3cm}
\epsfig{figure=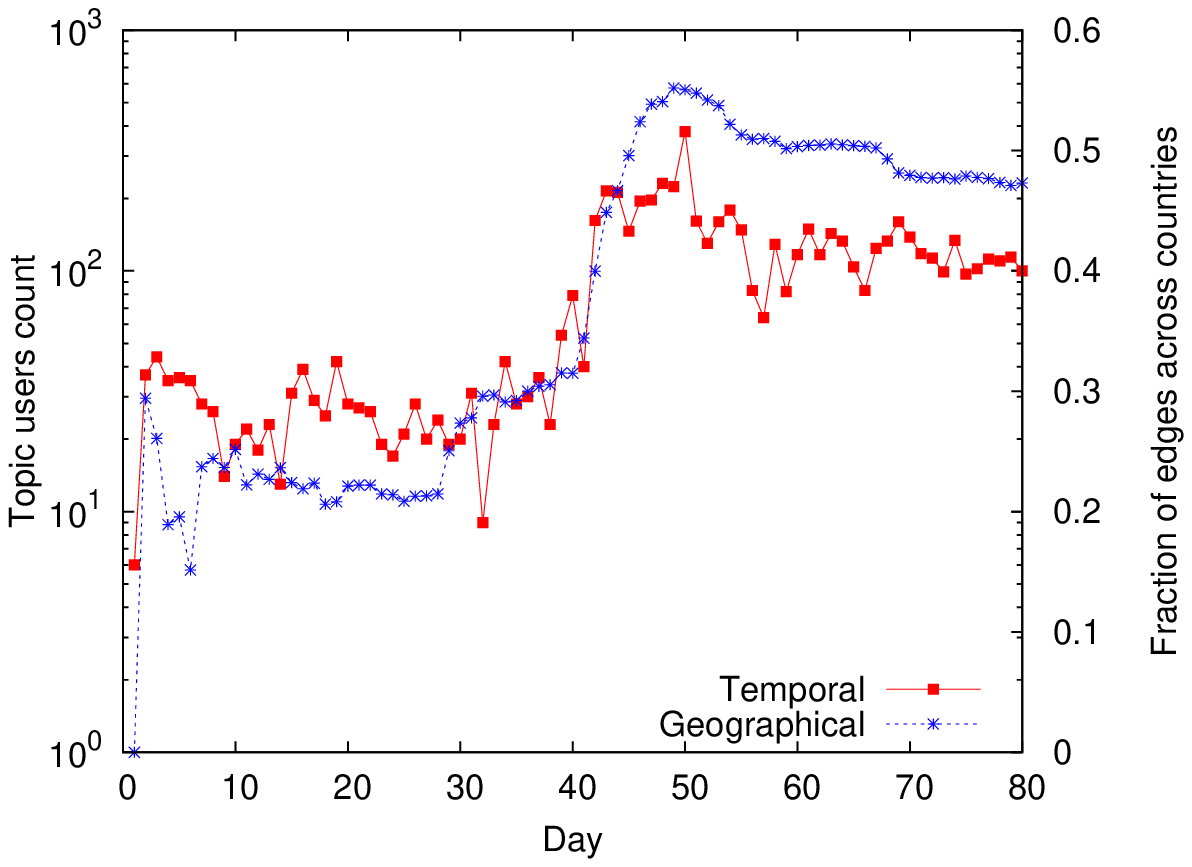,width=5.1cm,height=3.4cm}
\end{minipage} &
\begin{minipage}{5.3cm}
\epsfig{figure=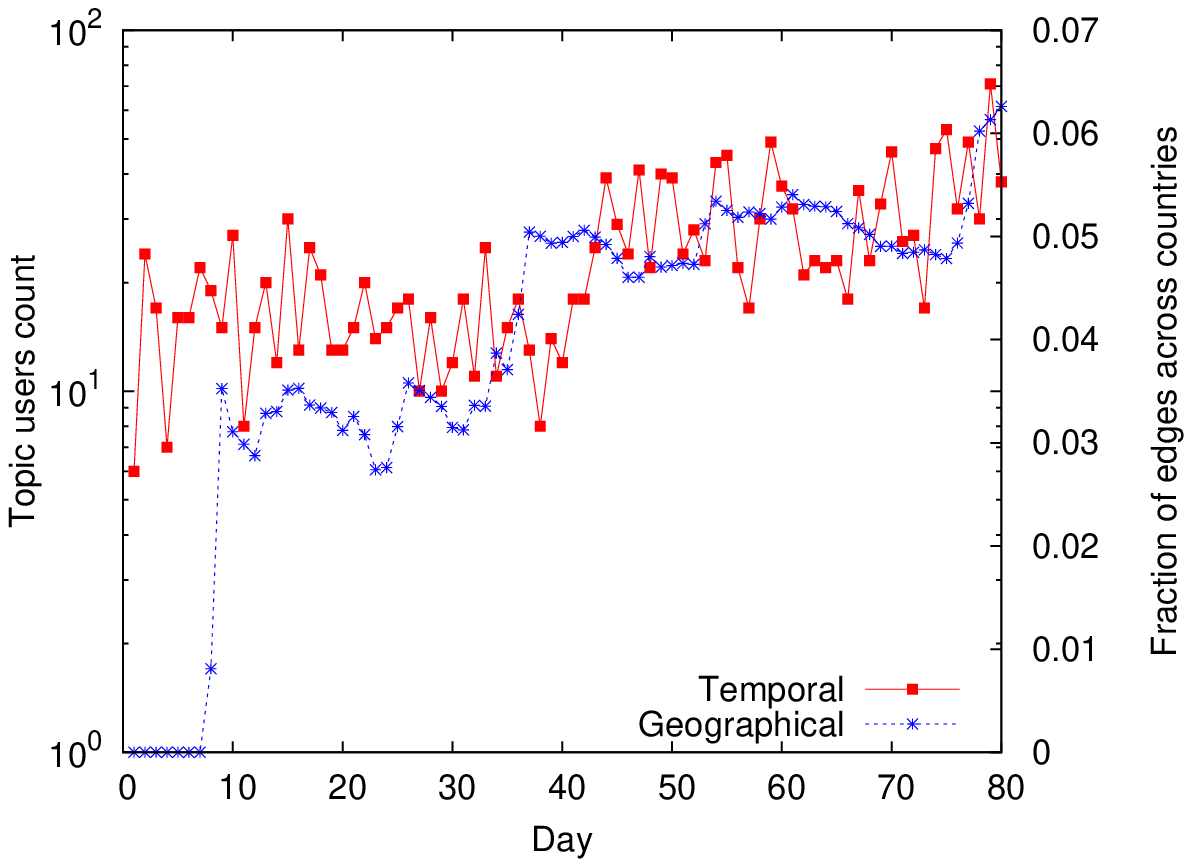,width=5.1cm,height=3.4cm}
\end{minipage} \\
(a) BARACK &(b) CAMBRIDGE &(c) HAMBURG\\
 \end{tabular}
\caption{Popularity vs edges crossing geography.}
\label{fig:GeoProperties}
\vspace{-.1in}
\end{figure*}

In order to establish the hypothesis, we investigated a geographical
property of the cumulative evolving graphs defined in
Section~\ref{sec:topology}. For each topic we determined the fraction
of edges in the cumulative evolving graph that went from one region to
another; that is, we studied the fraction of edges $(u \rightarrow v)$ such
that $u$ belongs to one region and $v$ is a user from another
region. The evolution of this fraction for three topics, one highly
popular, one with a medium level of popularity and one with a low
level of popularity (as defined in Section~\ref{sec:topology}) is shown
in Figure~\ref{fig:GeoProperties}.
We observe that the highly popular topic ``BARACK'' shows a high
fraction of edges crossing regional boundaries throughout its
evolution, ranging between 0.74 and 0.81. On the other hand the topic
with medium popularity, ``CAMBRIDGE'', has a low fraction of edges
crossing regions. It's noteworthy that an increase in the popularity
of the topic ``CAMBRIDGE'' is accompanied by an increase in the
fraction of edges crossing regional boundaries. This further supports
Hypothesis 3. The topic ``HAMBURG'' which has low popularity
shows a very small fraction of edges crossing regional boundaries.

\begin{figure}[htbp]
\centering
\subfigure[Mean]{
{\includegraphics[width=6cm,height=4cm]{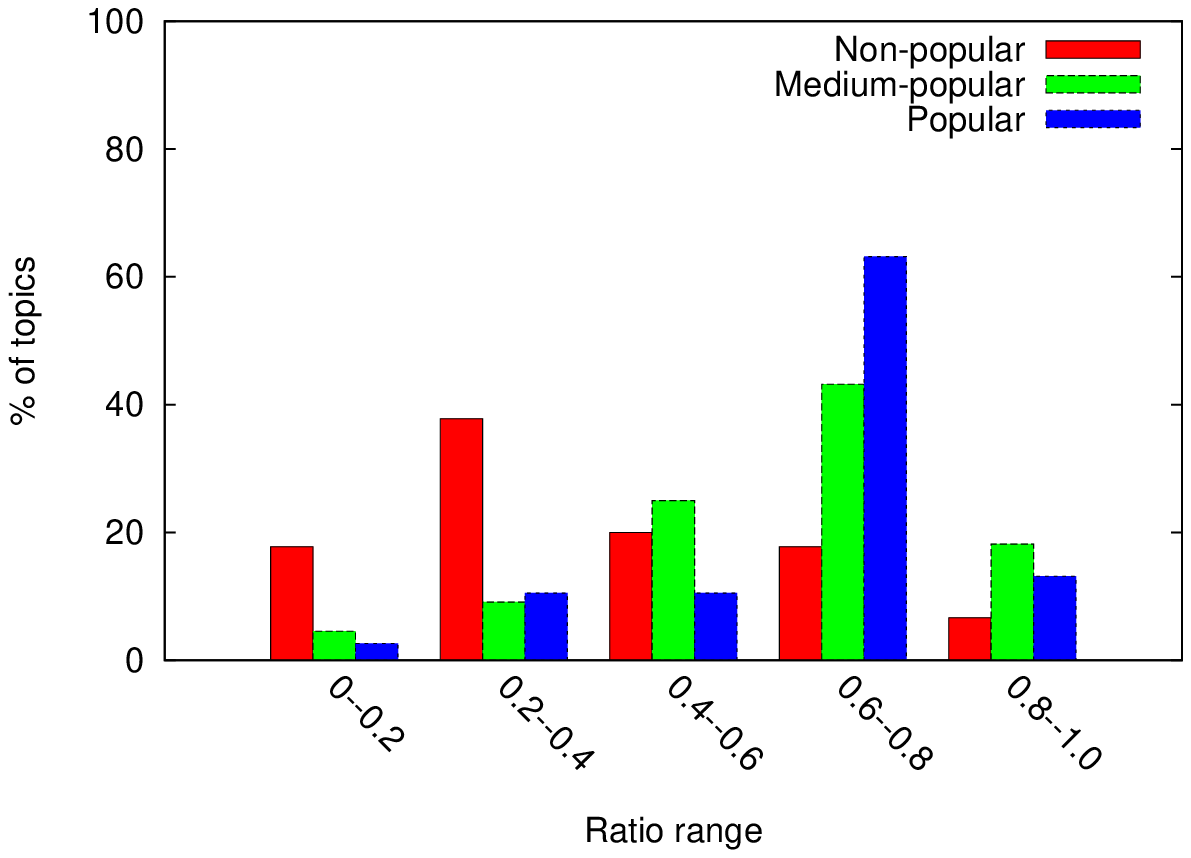}}
}
\subfigure[Median]{
{\includegraphics[width=6cm,height=4cm]{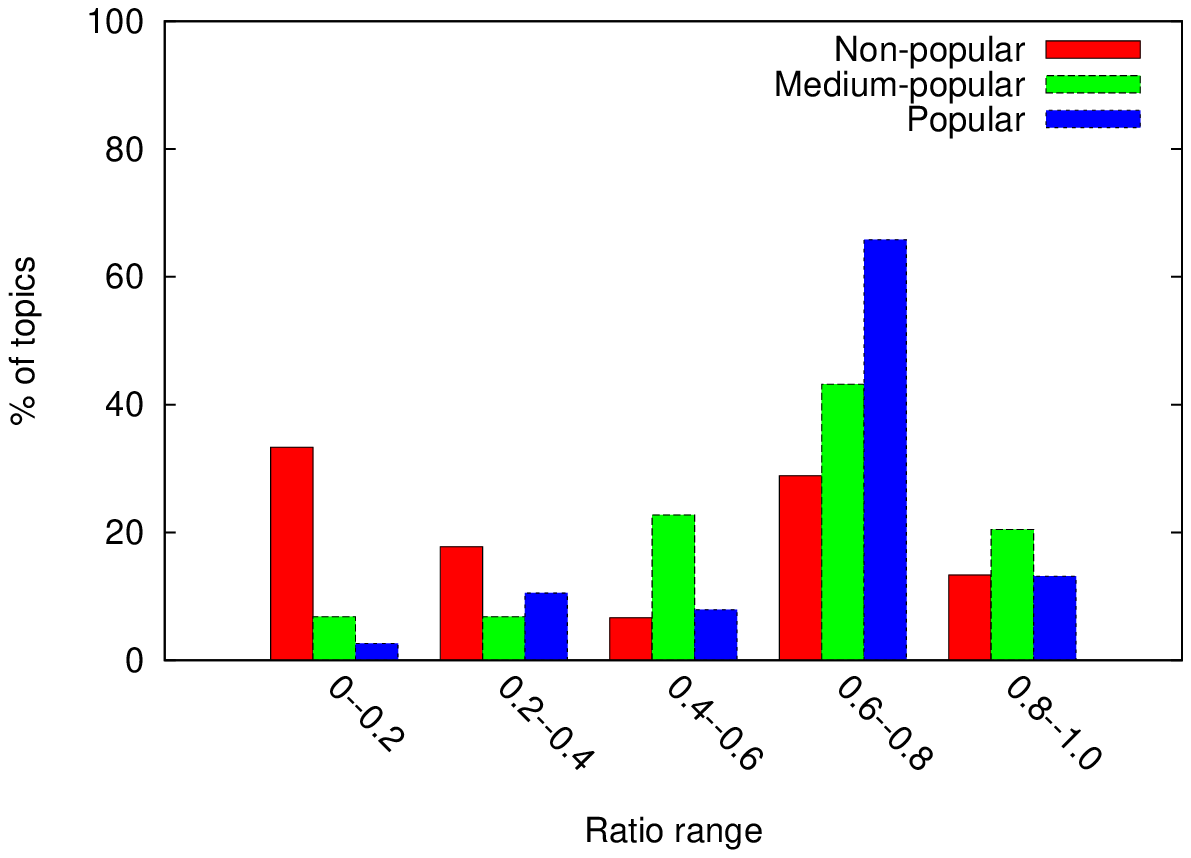}}
}
\caption{Fraction of edges cross regional boundaries.}
\label{fig:GeoMeanMedian}
\vspace{-.1in}
\end{figure}

To examine this phenomenon at an aggregate level we took 40 topics
from each category (as we had done in Section~\ref{sec:topology}) and
computed the mean and median of the fraction of edges crossing
regional boundaries for the entire period in our window where the
topic is tweeted on. We plotted a histogram using five different
ranges for this fraction (see Figure~\ref{fig:GeoMeanMedian}). 
This histogram clearly shows that the most popular topics tend to have
a very large fraction of edges crossing regional boundaries while the
least popular topics have cumulative graphs that generally evolve
within regional boundaries with small fractions of edges going to
other regions.

 \section{Concluding Remarks}
\label{sec:conclusions}

The studies we have presented in this paper have wide-ranging
implications, some of which, we hope, will be discovered in the
future. For now we present a brief discussion of those area we feel
our results may have an impact on.

Perhaps the most important implication pertains to the role and
impact of highly influential users (and consequently of highly
influential geographies). The rise of OSNs has been accompanied by a
triumphal narrative of democratization of communication through
technology, and while it is true that Twitter and other OSN platforms
have played an important role in giving voice to individuals who might
otherwise find it difficult to speak to an audience beyond their
immediate geography, our study shows that traditional holders of power
and influence have not been unseated.

Our hypothesis on how a giant component forms on Twitter--by the
merging of smaller tightly clustered sets of users--is an important
input into the sociology of how information is transacted on a social
network. There is reason to believe that despite the fact that OSN
platforms being the world closer, older notions of proximity and
community continue to contribute significantly to popularity in the
way described. Our study is broad in nature and captures a coarse
phenomenon that we hope will excite sociologist and invite them to
tease out the finer nuances that lie within such phenomena.

From an engineering standpoint issues of content distribution and
caching can be addressed from observing that highly popular topics
cross national boundaries. A closer study of which national boundaries
are crossed more often than others could underpin efficient content
placement methods.

Our results could also be of great interest to those involved in using
the vast reach of media like Twitter to advertise their products and
services. The notions of trust and reputation inherent in OSNs have
been leveraged to a great extent already for marketing purposes. Our
study could help advertisers and marketers figure out how best to use
these platforms for efficient and well-targeted marketing.

\end{document}